\documentclass[physrev,9pt,a4paper,reprint,superscriptaddress,amsmath]{revtex4-2}

\usepackage[charter]{mathdesign}
\DeclareSymbolFont{usualmathcal}{OMS}{cmsy}{m}{n}
\DeclareSymbolFontAlphabet{\mathcal}{usualmathcal}
\usepackage{graphicx}
\usepackage{dcolumn}
\usepackage{bm}
\usepackage{hyperref}
\hypersetup{  colorlinks,
              linktoc         = page,
              urlcolor        = {green!80!blue},
              linkcolor       = {orange!60!black},
              urlcolor        = {magenta!80!black}, 
              citecolor       = {magenta!80!black},
              anchorcolor     = {yellow}
}
\usepackage[dvipsnames]{xcolor}

\newcommand{\monthname}{%
  \ifcase\month
  \or January\or February\or March\or April\or May\or June\or July%
  \or August\or September\or October\or November\or December%
  \fi}

\newcommand{\bmsection}[1]{\par{\bf #1:}}

\begin{document}

\title{Optical lever for broadband detection of fluid interface fluctuations}

\author{Sreelekshmi C. Ajithkumar}
\affiliation{School of Physics and Astronomy, University of Nottingham, University Park, Nottingham, NG7 2RD, UK}

\author{Vitor S. Barroso}
\affiliation{School of Mathematical Sciences, University of Nottingham, University Park, Nottingham, NG7 2RD, UK}

\author{Patrik \v{S}van\v{c}ara}
\altaffiliation[Present address: ]{Institut N\'{e}el, CNRS, 25 avenue des Martyrs, 38042 Grenoble, France}
\affiliation{School of Mathematical Sciences, University of Nottingham, University Park, Nottingham, NG7 2RD, UK}

\author{Anthony J. Kent}
\affiliation{School of Physics and Astronomy, University of Nottingham, University Park, Nottingham, NG7 2RD, UK}

\author{Silke Weinfurtner}
\email{silke.weinfurtner@nottingham.ac.uk}
\affiliation{School of Mathematical Sciences, University of Nottingham, University Park, Nottingham, NG7 2RD, UK}
\affiliation{Centre for the Mathematics and Theoretical Physics of Quantum Non-Equilibrium Systems, University of Nottingham, Nottingham, NG7 2RD, UK}


\begin{abstract}
We exploit the optical lever principle to detect minute fluctuations of a liquid-air interface. Waves propagating on the interface deflect a specularly reflected laser beam, inducing angular deviations captured by a dual-element photodiode. We implement this principle in a compact set-up that includes a temperature-controlled fluid sample. This allows us to detect deflection angle fluctuations across five orders of magnitude in frequency, from individual low-frequency surface eigenmodes to the thermal distribution of high-frequency capillary waves. In addition to demonstrating the method's versatility and broad dynamical range, we highlight practical considerations in characterising liquid interface dynamics, bridging established optical methods with their application to fluid and soft-matter systems.
\end{abstract}

\maketitle
When light encounters an optical interface, it may be partially reflected, absorbed, or transmitted, with details depending on the properties of optical media and the angle of incidence. Each of these interactions carries information about the interface. Here, we explore how a reflected beam can serve as a sensor for detecting waves propagating on a liquid-gas or liquid-liquid interface. Such interface excitations may arise from inherent thermal fluctuations or external mechanical vibrations and cover a wide frequency range, spanning from Hz up to GHz.

Interferometry is a powerful tool for detecting changes and fluctuations in fluid interfaces from phase variations of transmitted or reflected light. With standard set-ups, such as Michelson or Mach-Zehnder interferometers, a liquid sample is placed along the path of a probe laser beam, which is typically recombined -- and interfered -- with a reference. The resulting interference fringes are then tracked over time to recover local variations on the sample surface with sub-nanometre precision~\cite{westlake1967fluid, mitsui2013measurements,verma2015universal,slavchov2021characterization}. These interferometric measurements allowed the non-invasive characterisation of surface waves across a wide range of frequencies to ultimately determine the rheological or physiochemical properties of liquids~\cite{behroozi2003noninvasive,behroozi2010stokes,tachizaki2006scanning, chaudhary2021universal}. 

Under the same principles, digital holography can be used to reconstruct entire regions of a fluid surface over time from interference images formed from the sample~\cite{barroso2023digital, shmyrov2019capillary}. Holographic measurements offer both spatial and temporal spectral characterisation of surface waves and, hence, a complete picture of their dispersion relation. However, the frequency range typically accessible in digital holography is limited by the sampling rate restrictions of digital cameras used in the set-up.

The detection of scattered back-reflected light offers an alternative approach. Early light scattering experiments by Katyl and Ingard \cite{katyl1967line,katyl1968scattering} revealed spectral broadening in light reflected from the interfaces of methanol and isopropanol using a Fabry-Pérot interferometer. Bouchiat et al. \cite{bouchiat1968demonstration} later extended this technique to ethyl ether and glycerol, attributing modifications in the scattered light's spectrum to short-wavelength, thermally excited waves propagating on the interface. The fundamental relation between fluctuations and dissipation \cite{kubo1966fluctuation} enabled the use of this thermal interface noise as a minimally-invasive probe of surface tension of liquid-gas interfaces. Subsequent experiments by Huang and Webb \cite{huang1969viscous}, as well as Meunier \cite{meunier1969diffusion} determined the critical exponent for surface tension as a function of temperature near the critical point, initiating a series of studies that employed light scattering to measure both the surface tension and viscosity in various fluids \cite{sohl1978novel}. This approach led to further investigations into the rheological properties of soft matter, including polymers \cite{huang1998effects,cicuta2004recent}, the freezing of capillary waves on oligomers \cite{streit2008cooling}, and the propagation of surface waves on soft gels \cite{kikuchi1994complex,keunho2001thickness}. For a comprehensive review, see \cite{langevin2021light}.

A specularly reflected beam equally serves as a local sensor of interface fluctuations, operating on the principle of optical lever. The fluctuating interface alters the direction of the reflected beam. If the liquid surface tilts by a small angle $\theta$, the beam undergoes a corresponding directional shift, translating angular deviations into lateral displacements that can be precisely detected using position-sensitive photosensors, such as multiple-element photodiodes or pixel detectors, providing a direct means for measuring inclination angles on the order of nanoradians. This principle led to the development of \emph{surface fluctuation specular reflection spectroscopy} (SFSRS) as an alternative to light scattering techniques \cite{tay2008probing}. With the availability of fast analog-to-digital converters, SFSRS enables broadband, real-time fluctuation measurements, making it a valuable tool for fully non-invasive viscosimetry \cite{pottier2011high}.

One of the key advantages of SFSRS over scattering-based techniques is its substantially lower power requirement. Like earlier approaches, it provides the sensitivity needed to detect thermal noise across a broad range of liquid interfaces \cite{aoki2018thermal}, proving instrumental in studies of boundary conditions in liquid thin films \cite{pottier2015boundary} and, more recently, in monitoring ice melting \cite{mitsui2019fluctuation}. Compared with point-like interferometric methods, which can achieve sub-nanometre sensitivity, SFSRS offers complementary benefits: it is mechanically simpler, less sensitive to environmental drifts such as air currents along the free beam path, and directly sensitive to surface gradients rather than absolute height fluctuations. To further improve its signal-to-noise ratio, more elaborate experimental set-ups have been developed, including a two-beam approach \cite{raudsepp2012two}, where deflections of two independent laser beams are correlated, and single-beam correlation measurements that surpass the shot-noise limit of photodetectors \cite{mitsui2013measurements}.

In this work, we implement a simplified SFSRS set-up to study broadband interface dynamics. The bandwidth of the readout circuit, limited to about 300~kHz, is sufficient to resolve both discrete surface modes and the quasi-continuous spectrum of thermally driven interface fluctuations at higher frequencies. These complementary regimes are sensitive to the fluid's physiochemical properties, which vary with temperature and therefore require precise thermal control. To ensure stable conditions, we developed a temperature-controlled sample cell with high thermal conductivity and partial top coverage, which minimises evaporation at elevated temperatures and enables longer acquisition times. While the well understood dynamics of thermal noise on liquid interfaces allows us to benchmark the performance of SFSRS under controlled experimental conditions, the simultaneous detection of discrete, mechanically driven surface modes within the same set-up has not previously been achieved.

\section{Experimental set-up}

\subsection{Principle}
\begin{figure}[htbp]
    \centering
    \includegraphics[scale=1]{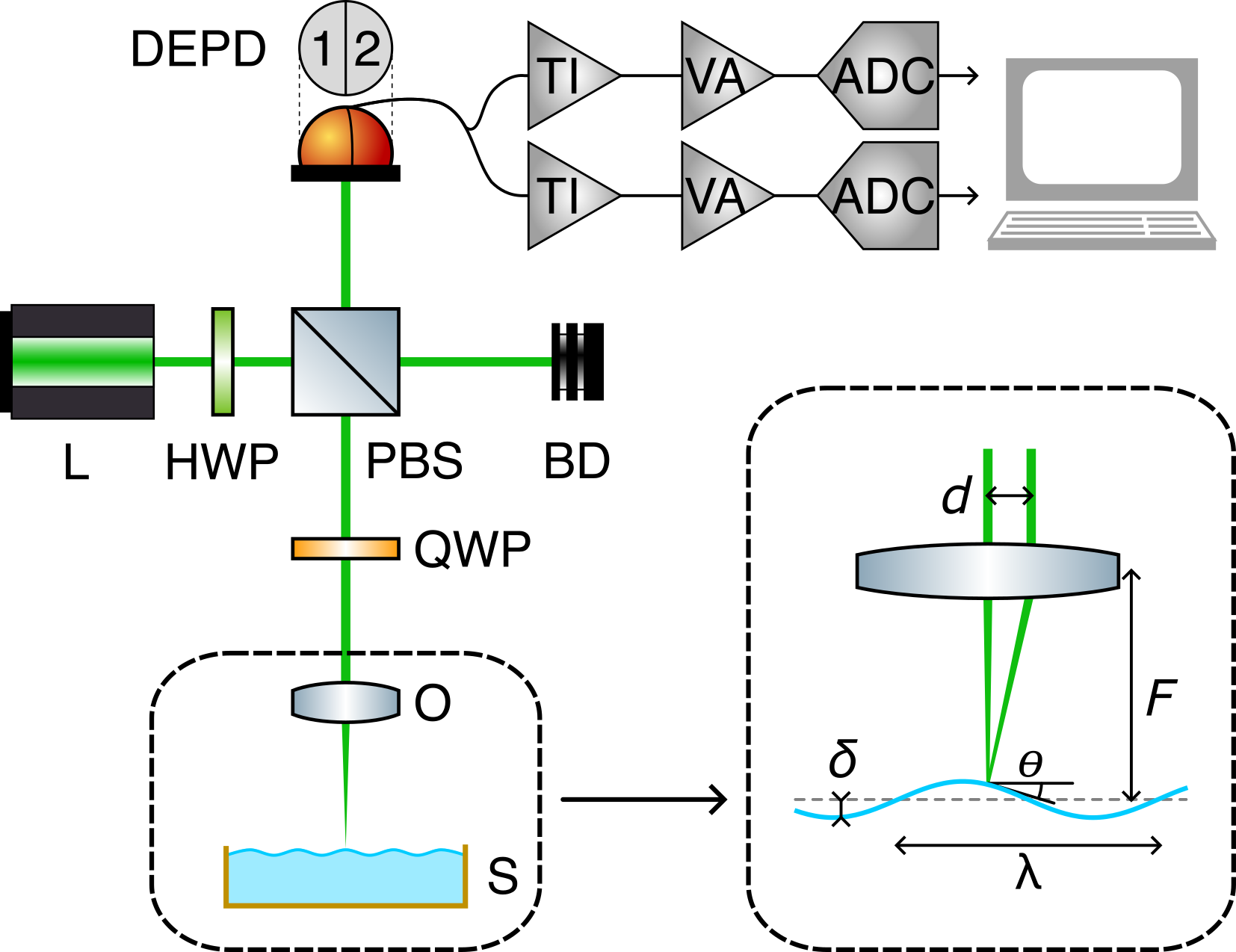}
    \caption{Scheme of the experimental set-up. L, laser; HWP, half-wave plate; PBS, polarising beam splitter; BD, beam dump; QWP, quarter-wave plate; O, objective lens; S, fluid sample; DEPD, dual-element photodiode; TI, transimpedance amplifier; VA, voltage amplifier; ADC, analog-to-digital converter. Inset details the geometry of the inclined interface. A surface wave with wavelength $\lambda$ and small amplitude $\delta$ induces a local inclination angle $\theta$. Due to this inclination, the reflected beam acquires a displacement $d$ with respect to the incident beam in the objective plane. $F$ denotes the focal length of the objective lens.}
    \label{fig:1}
\end{figure}

The optical arrangement for SFSRS is outlined in Fig.~\ref{fig:1}. A continuous laser beam (L, M Squared Equinox, wavelength $\Lambda = 532$~nm and power approx. $1$ mW) is reflected by a polarising beam splitter (PBS) towards the objective lens (O) and the fluid sample (S). The polarisation of the beam reflected off the sample is such that it transmits through the PBS and illuminates a dual-element photodiode (DEPD). The photocurrents of both elements are independently amplified and converted to voltage by a transimpedance amplifier (TIA, Thorlabs AMP140, conversion ratio $R_f = 10$~kV/A), further amplified by a custom voltage amplifier (VA, gain $G = 10$), digitised by an analogue-to-digital converter (ADC, Teledyne ADQ14 14-bit), and eventually saved for further processing. We note that this approach differs fundamentally from traditional balanced photodetection schemes, which are designed to suppress common-mode amplitude noise of the laser field and thereby improve the signal-to-noise ratio in precision optical measurements \cite{hobbs1997ultrasensitive}.

The interface of the fluid sample coincides with the focal plane of the objective lens (O, focal length $F$). Focusing of the Gaussian beam by the objective results in a probing spot with the $1/e^2$ intensity radius $w_0$,
\begin{equation}
    w_0 = \frac{\Lambda}{\pi\mathrm{NA}}\,,
    \label{eq:spotsize}
\end{equation}
where $\mathrm{NA} = w/F$ defines the effective numerical aperture determined by the beam radius $w$ at the entrance of the objective with focal length $F$.

The inset in Fig.~\ref{fig:1} illustrates the scenario where a normally incident beam reflects off the surface wave with wavelength $\lambda$ and amplitude $\delta$. The angular deflection of the reflected beam is determined by the local inclination angle $\theta$, as shown in the figure. We assume $\delta \ll \lambda$, which means $\theta \ll 1$~rad. The objective translates the beam's angular deflection into a lateral displacement $d = 2F\theta$. Such a beam unevenly illuminates the DEPD, and the displacement $d$ is detected as a difference in the independently measured voltage signals $V_1$ and $V_2$, acquired by our readout line and proportional to the photocurrents induced on the elements. The inclination angle $\theta$ can be then calculated as (see Supplemental document)
\begin{equation}
    \theta = \frac{\pi \mathrm{NA}}{8} \cdot \frac{V_1 - V_2}{V_1 + V_2}\,.
    \label{eq:exptheta}
\end{equation}
In practice, we intentionally offset the photodiode so that the signal difference fluctuates around a nonzero mean. This procedure, detailed in Supplemental document, suppresses artefacts arising from the finite gap between the DEPD elements and affects only the DC component of the calculated surface inclination spectra.

We finally note that the angular sensitivity of the optical lever depends on the size of the probing spot $w_0$. A smaller spot produces a wider diffraction pattern, thereby reducing sensitivity to small fluctuations of $\theta$. However, a sufficiently small probing spot is required to resolve short-wavelength surface modes studied below. In our set-up, the input beam is typically focused to $w_0 = 13.3~\mathrm{\mu m}$, and the total optical power reaching the DEPD is about $30~\mathrm{\mu W}$. This corresponds to a flat shot-noise floor in the measured power spectral density of $\theta$ of approximately $2\times10^{-18}~\mathrm{rad^2/Hz}$ (see Supplemental document). In practice, however, the sensitivity is limited mainly by mechanical vibrations and electronic noise of the readout circuit rather than photon statistics, as discussed later in the text.

\subsection{Sample cell}

\begin{figure}[htbp]
    \centering
    \includegraphics[scale=1]{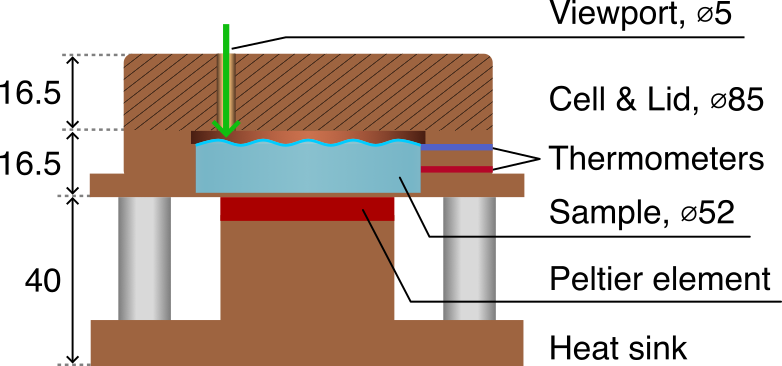}
    \caption{Temperature-controlled experimental cell. Dimensions are given in millimetres.}
    \label{fig:2}
\end{figure}

The sample cell is machined from copper, and its cross section is schematically shown in Fig.~\ref{fig:2}. The sample volume is a cylindrical cavity with diameter $D = 52$~mm and depth $H = 12$~mm. These dimensions are chosen to accommodate a standard $2^{\prime\prime}$ mirror (Thorlabs BB2-EO2), whose reflective surface coincides with the height of a sharp $90^\circ$ edge that serves to pin the liquid-air interface. Temporarily replacing the liquid sample with the mirror (and compensating for increased reflectivity of the mirror by means of a neutral density filter placed between the laser source and beam splitter) allows to focus the objective lens to the correct height (see \cite{sree2025thesis} for a discussion on imperfect focusing) and acquire ``dummy'' data sets that distinguish hydrodynamically-induced signals from the mechanical vibrations of the optical components as discussed below. To minimise external disturbances, the set-up is mounted on a passively isolated optical table. The temperature of the sample cell is controlled with accuracy on the order of few mK via a custom proportional-integral-derivative (PID) controller (see Supplemental document for a detailed description of this system and its performance). To minimise evaporation and improve the overall thermal stability of the sample, the cell can be closed by a thick copper lid with several 5-mm viewports, allowing the laser beam to reflect off the liquid interface and exit the cell. Finally, multiple reflections of the probe beam between the liquid interface and the bottom of the sample cell are suppressed by coating the bottom of the sample cell with matte black paint.

\subsection{Surface waves}\label{sec:surf_waves}
The fluctuating interface height can be decomposed in terms of normal modes in time and space. In cylindrical geometry, with radius $r$ and polar coordinate $\phi$, a normal mode with frequency $\omega$ will have a corresponding spatial profile $\psi_{mn}(r,\phi)$, which can be written as
\begin{equation}
    \psi_{mn} (r,\phi) = R_{mn}(r) \exp(im\phi)\,,
    \label{eq:psi}
\end{equation}
where the integer $m$ is referred to as the azimuthal number, and the radial component $R_{mn}(r)$ satisfies Bessel's equation of order $m$. Here, the non-negative integer $n$ labels the various radial profiles $R_{mn}$, each related to a different normal mode with frequency $\omega$. The properties of the contact line between the liquid interface and the sample cell walls will ultimately determine the boundary conditions for the radial components $R_{mn}$. When the interface sits perfectly orthogonal to the walls and the contact line is completely free to move, $R_{mn}(r) \propto J_{|m|}(k_{mn}r)$, where $J_{|m|}$ is the $m$-th order Bessel function of the first kind, and $k_{mn}$ denotes the corresponding wavenumber. The latter is determined by the boundary condition on the cell wall $r_0 \equiv D/2$, which, for the freely-moving interface, is a Neumann (full-slip) boundary condition, i.e. $J^\prime_{|m|}(k_{mn}r_0) = 0$. Conversely, our sample cell pins the interface to the wall, and hence, the contact line does not move. In this case, the radial profiles satisfy Dirichlet (fixed) boundary conditions, for which $J_{|m|}(k_{mn}r_0) = 0$.

The wavenumbers $k$ and frequencies $\omega = 2\pi f$ of surface waves are tied together by a dispersion relation, which reads for a liquid-air interface,
\begin{equation}
  \omega^2 = \left(gk + \frac{\sigma}{\rho} k^3\right) \tanh\left(h_0 k\right)\,,
  \label{eq:disp}
\end{equation}
where $g$ is the gravitational acceleration, $\sigma$ the surface tension, $\rho$ the density of the liquid, and $h_0$ the depth of the unperturbed liquid volume. In cases like ours, where the contact line is fixed, the frequency of surface normal modes is known to deviate from the dispersion relation~\eqref{eq:disp} when computed with wavenumbers given exactly by Dirichlet boundary conditions~\cite{Henderson1994Surface-waveLine,coccairo1993experimental,picard2007resonance,howell2000measurements,kidambi2009capillary,kidambi2009meniscus}. This deviation is particularly relevant for the lowest available wavenumbers $k_{mn}$, which are coarsely spaced (highly discretised), appearing as mechanically-driven, low-frequency ($<100~$Hz) waves. At higher frequencies, surface waves typically have short wavelengths (large wavenumbers), $\lambda\lesssim \sqrt{\sigma/(\rho g)}$ and $\lambda \ll h_0$, and their dispersion relation~\eqref{eq:disp} is dominated by surface tension, i.e. $\omega^2\approx\omega_c^2=(\sigma/\rho)~k^{3}$. At these scales, boundary effects are negligible, and these capillary waves display an effectively continuous distribution of wavenumbers. As we show below, our set-up is capable of probing both regimes, including the spectral crossover from discrete to quasi-continuous wave behaviour.

\section{Results}

\subsection{High-frequency thermal noise}
The detection of the continuous spectrum of thermally induced capillary waves constitutes a benchmark of our approach. We employ Eq.~\eqref{eq:exptheta} to acquire time-resolved fluctuations of surface inclination and calculate their frequency spectrum, which is qualitatively equivalent to the spectrum of height fluctuations. To obtain quantitative height information, our method must be calibrated, e.g. through interferometric measurements~\cite{barroso2023digital}.

The range of wavenumbers accessible in our experiment is inherently limited. The smallest wavenumber (corresponding to the longest wavelength) supported by the sample cell is $k_1 \sim 2\pi/D$, where $D$ is the cell diameter. Although the spectrum is discrete at these scales (cf. Sec.~\ref{sec:surf_waves}), it is approximated here as continuous by extending the integration down to $k = 0$. At the other end of the spectrum, the detection of short-wavelength fluctuations is constrained by the finite size of the probe beam, which imposes an effective upper cut-off at $k_2 \sim 2\pi/(2w_0)$. These constraints are incorporated into the measured inclination spectrum~\cite{aoki2012spectral}, expressed as
\begin{equation}
    S(f) = \int_{0}^{\infty} k^3 \, e^{-w_0^2k^2/2} \, P(k,\omega) \,dk\,,
    \label{eq:sf}
\end{equation}
where $P(k,\omega)$ denotes the full thermal spectral function of height fluctuations. The latter can be derived under the assumption of a finite-temperature liquid-gas interface~\cite{bouchiat1971spectre}. By neglecting the influence of the gaseous phase, the function reads
\begin{equation}
    P(k,\omega) = \frac{k_B Tku^2}{\pi \rho \omega^3} \mathrm{Im}\left\{\left[(1-iu)^2+y-\sqrt{1-2iu}\right]^{-1}\right\}\,,
    \label{eq:pkomega}
\end{equation}
where $k_B$ is the Boltzmann constant, $T$ denotes absolute temperature, and the auxiliary functions $u$ and $y$ are defined in terms of the surface-wave viscous dissipation rate, $\gamma=2(\mu/\rho) k^2$, and the capillary dispersion frequency, $\omega_c = \sqrt{\sigma/\rho}~k^{3/2}$, as
\begin{equation}
    u =\frac{\omega}{\gamma}= \frac{\rho\omega}{2 \mu k^2}\,,\qquad
    y=\left(\frac{\omega_c}{\gamma}\right)^2=\frac{\rho \sigma}{4\mu^2 k}\,,
\end{equation}
where $\mu$ denotes the liquid's dynamic viscosity.

\begin{figure}[htbp]
    \centering
    \includegraphics[scale=1]{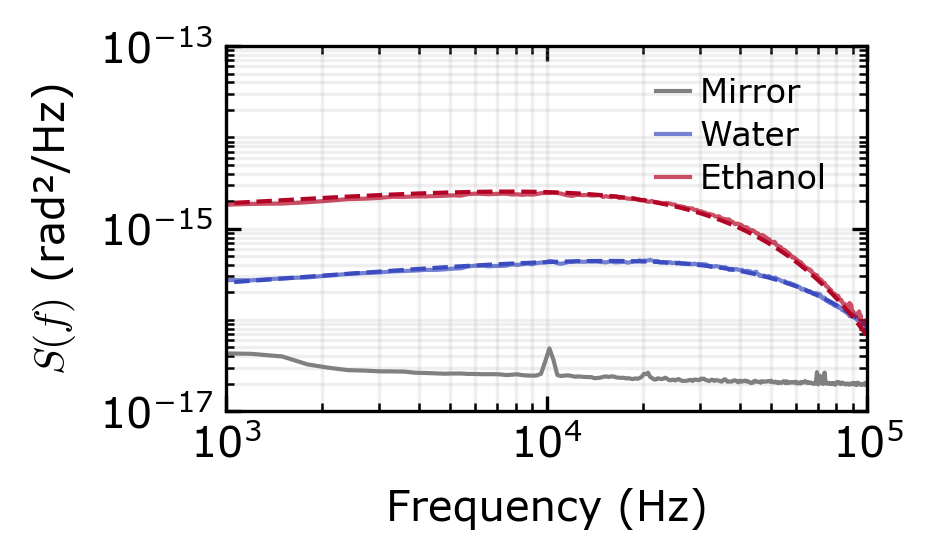}
    \caption{Surface inclination spectra of chromatography water (blue line) and reagent-grade ethanol (red line), measured at $20^\circ$C using SFSRS. The dashed lines overlaying experimental data are theoretical curves of thermal noise spectra with no free parameters. Grey line, measured with a mirror placed inside the sample cell, characterises the vibrational noise floor. In this series of experiments, $\mathrm{NA} = (12.7\pm 0.3)\times 10^{-3}$.}
    \label{fig:3}
\end{figure}

In Fig.~\ref{fig:3}, we compare the power spectral density of surface inclination angle (\emph{surface inclination spectrum}) measured by SFSRS at $20^\circ$C for chromatography water (solid blue line) and reagent-grade ethanol (solid red line), with thermal spectra calculated from Eqs.~\eqref{eq:sf} and \eqref{eq:pkomega}, shown as overlapping dashed lines. The theoretical and experimental spectra exhibit close agreement over the range from approximately 1 to 100~kHz. Notably, this match is achieved without any fitting parameters, using only tabulated fluid properties (density, surface tension, viscosity) summarised in Table~\ref{tab:properties}.

Above the $100$-kHz mark, SFSRS becomes insensitive to the extremely short wavelength of capillary waves (cf. the upper cut-off wavenumber $k_2$). We can hence estimate the maximum accessible frequency as
\begin{equation}
    f_2 = \frac{\omega_c(k_2)}{2\pi} = \frac{\pi^2}{2} \sqrt{\frac{\sigma}{\rho}\left(\frac{\mathrm{NA}}{\Lambda}\right)^3}\,.
    \label{eq:fmax}
\end{equation}
For the employed numerical aperture $\mathrm{NA} = (12.7\pm 0.3)\times 10^{-3}$, we obtain the maximum frequency of approx. 151~kHz for water and 95~kHz for ethanol, which justifies the considered frequency range.

\begin{table}[htbp]
\caption{Physiochemical properties of water \cite{water-data} and ethanol \cite{ethanol-data}.}
  \label{tab:properties}
  \centering
\begin{tabular}{lcccc}
\hline
Fluid & $T$ ($^\circ$C) & $\rho$ (kg/m$^3$) & $\sigma$ (mN/m) & $\mu$ ($\mu$Pa.s) \\
\hline
    Water   & 20 & $998.2$ & $72.74$ & $1001.6$ \\
            & 30 & $995.6$ & $71.19$ & $797.2$  \\
            & 40 & $992.2$ & $69.60$ & $652.7$  \\
    Ethanol & 20 & $791.2$ & $22.36$ & $1203$   \\
\hline
\end{tabular}
\end{table}

We now compare our data with the noise floor. The most significant noise source in this case is the TIA's input current noise (see Supplemental document for a detailed analysis of noise sources), which likely limits our measurements beyond the considered frequency range, 1-100~kHz; accessing higher frequencies is further limited at approx. 300~kHz by the bandwidth of the voltage amplifier. However, mechanical vibrations of the set-up may couple with the differential photocurrent and induce apparent inclination signals through Eq.~\eqref{eq:exptheta} in the present frequency range. This purely mechanical signal can be directly measured by replacing the fluid sample with a mirror, as already outlined. The corresponding spectrum (grey line) is the result of relative motion of the DEPD in its mount with respect to the experimental cell, and the amplitude of this signal is at least one order of magnitude lower than the signals measured in the fluid. Low level of ambient noise measured in-situ, combined with the close match of experimental and theoretical curves, indicate successful measurements of thermal noise on the surface of water and ethanol. Curiously, the measured spectral amplitude is about five times lower than the estimated total noise floor ($9.8\times10^{-17}~\mathrm{rad^2/Hz}$; see Supplemental document), indicating that the electronic noise of the readout circuit is lower than initially expected.

\begin{figure}[htbp]
    \centering
    \includegraphics[scale=1]{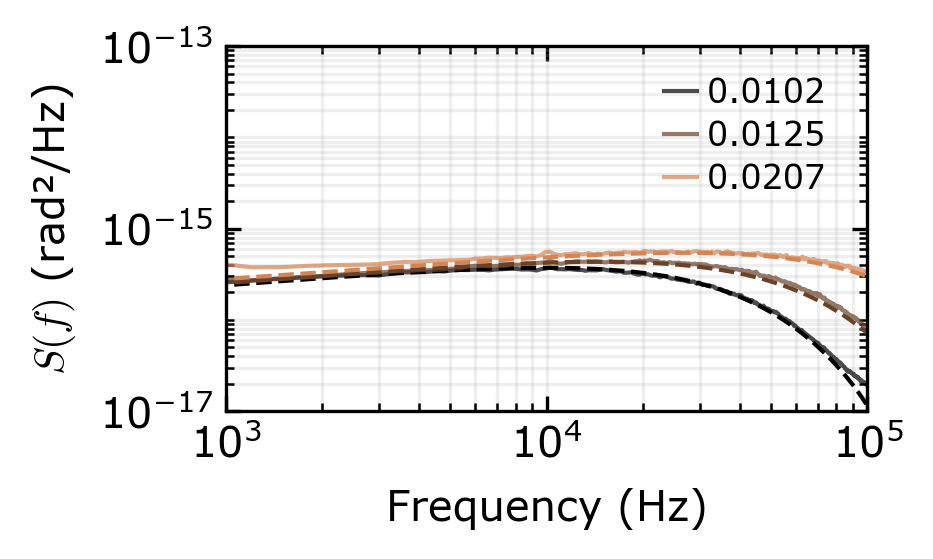}
    \caption{Surface inclination spectra of water at $20^\circ$C, measured for different numerical apertures, as specified in the legend. Solid lines represent experimental data, while dashed lines indicate the corresponding theoretical predictions.}
    \label{fig:4}
\end{figure}

Although the spectra measured in water and ethanol are obtained at the same temperature, the amplitude of thermal noise and the overall shape of the spectral function differs between them. In Fig.~\ref{fig:3}, this is given by their different physiochemical properties, with the dominant contribution of surface tension, which is about 3 times lower for ethanol. However, the shape of the spectral function can also be influenced by the numerical aperture of the objective. The objective sets the beam's spot size (Eq.~\eqref{eq:spotsize}) and consequently controls the upper cut-off frequency $f_2$ (Eq.~\eqref{eq:fmax}). We demonstrate this effect in Fig.~\ref{fig:4}. Each data series corresponds to the same system, i.e. thermal noise of water kept at $20^\circ$C, with differences in the focal length of the objective ($50$, $45$ and $35$~mm) and the diameter of the input beam ($2w$ between $1.25$ and $1.45$~mm), resulting in different $\mathrm{NA}$ values specified in the legend. Increasing $\mathrm{NA}$ results in a larger $f_2$, and the roll-off of the spectrum takes place at higher frequencies than previously discussed. On the other hand, increasing the $\mathrm{NA}$ and the accessible spectral range leads to an increased level of noise (see Supplemental document, Eq.~(S11)). In the practical application of SFSRS, one must hence consider a trade-off between the signal and noise amplification.

\begin{figure}[htbp]
    \centering
    \includegraphics[width=70mm]{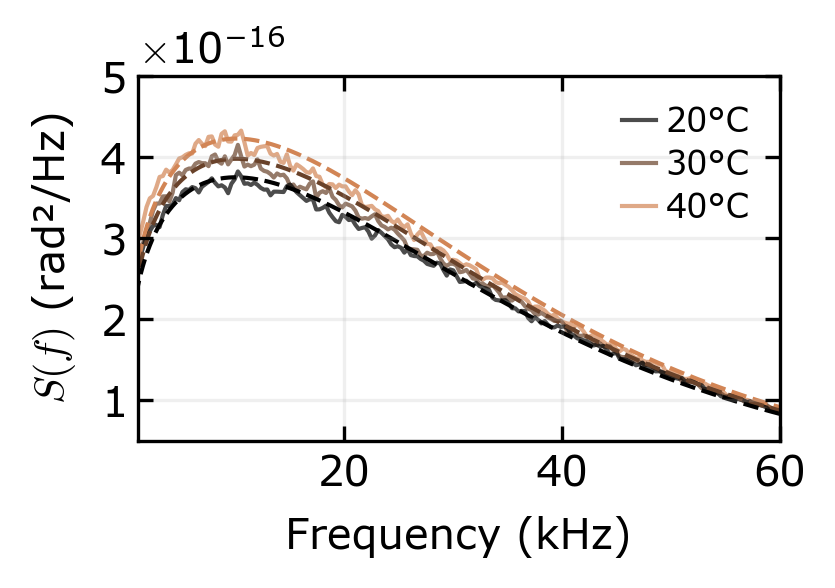}
    \caption{Surface inclination spectra measured in water, for different temperatures, as specified in the legend. The temperature is kept constant down to 5-mK fluctuations over the course of experiments (see Supplemental document for details). Consistently with previous figures, solid lines denote experimental data and dashed lines represent theoretical spectra. In this series of experiments, $\mathrm{NA} = (10.4\pm0.3)\times10^{-3}$.}
    \label{fig:5}
\end{figure}

To demonstrate the method's ability to discern thermal noise of the \emph{same} liquid kept at \emph{different} temperatures, we plot in Fig.~\ref{fig:5} inclination spectra of water kept at 20, 30 and $40^\circ$C, as solid lines; dashed lines again denote relevant theoretical predictions. A measurable difference is visible especially around 10~kHz, near the temperature-dependent maximum of the spectral function. This shows the ability of SFSRS to perform minimally-invasive thermometry, which becomes useful in applications where the presence of conventional thermometers in impractical, e.g. in thin fluid films. Note that temperature enters Eqs.~\eqref{eq:sf} and \eqref{eq:pkomega} explicitly, but also via temperature-dependent fluid properties. These parameters can be tuned, e.g. by adding surfactants or thickening agents \cite{Takamura2012}, to engineer a desired thermal spectral profile.

\subsection{Low-frequency modes}
\begin{figure}[htbp]
    \centering
    \includegraphics[scale=1]{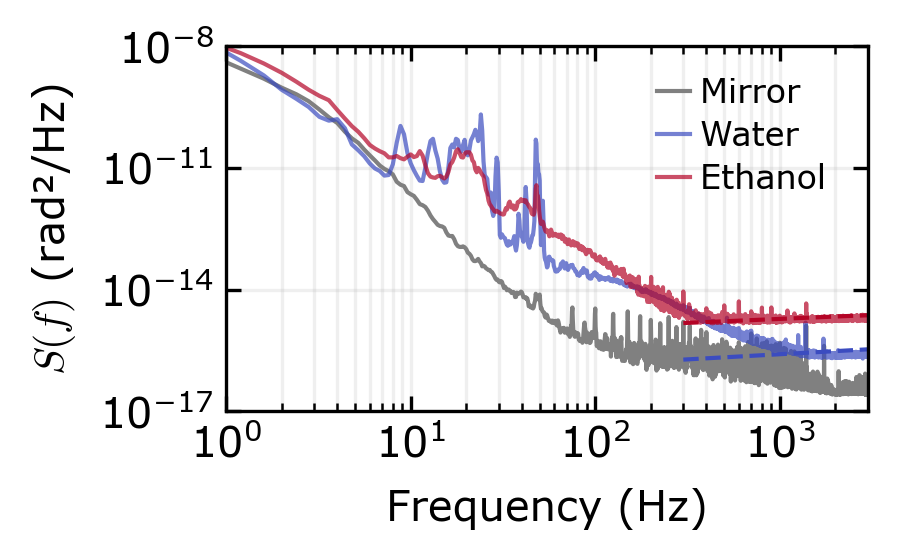}
    \caption{Low-frequency surface inclination spectra. As in previous figures, solid lines denote experimental data and dashed lines the theoretical model. Distinct peaks in the spectra of water and ethanol correspond to individual surface modes, excited on the interface due to ambient, mainly mechanical, noise. In this series of experiments, $\mathrm{NA} = (12.7\pm0.3)\times10^{-3}$.
    }
    \label{fig:6}
\end{figure}

As previously discussed, high density of modes oscillating at frequencies above 1~kHz allows treating the thermally-induced surface noise by the means of a continuous spectral function. This behaviour changes at low frequencies, where the density of eigenmodes reduces, opening the possibility of individually resolving them. In Fig.~\ref{fig:6}, we examine the low-frequency inclination spectrum, which is accessible in our set-up be reducing the ADC's acquisition rate and extending the acquisition time. The spectra are again obtained by Fourier transforming the differential DEPD signal. The control experiment using a mirror (grey line) displays an approximately $1/f^2$ background between 1 and 100~Hz, attributed to the finite acquisition time, followed by a series of peaks that are multiples of the line frequency, 50~Hz.

Let us now compare the control data set (grey line) with analogous spectra measured in water and ethanol, both at $20^\circ$C (blue and red lines). Between approx. 10 and 100~Hz, the spectra display a series of peaks, whose amplitudes are considerably higher than the baseline signal of the mirror, and of the predicted thermal noise (dashed lines; these match the experimental data for $f > 500$~Hz). The peaks therefore correspond to specific surface modes that are excited by the ambient mechanical noise. Their position is determined through Eq.~\eqref{eq:psi}, for some values of integers $m$ and $n$, and the corresponding boundary condition. As we discuss below, the fluid's behaviour at the boundary depends on wave amplitude, which is given by the nature of stochastic driving and frequency-dependent damping. In our case, these factors are not fully characterised.

\begin{figure}[htbp]
    \centering
    \includegraphics[scale=1]{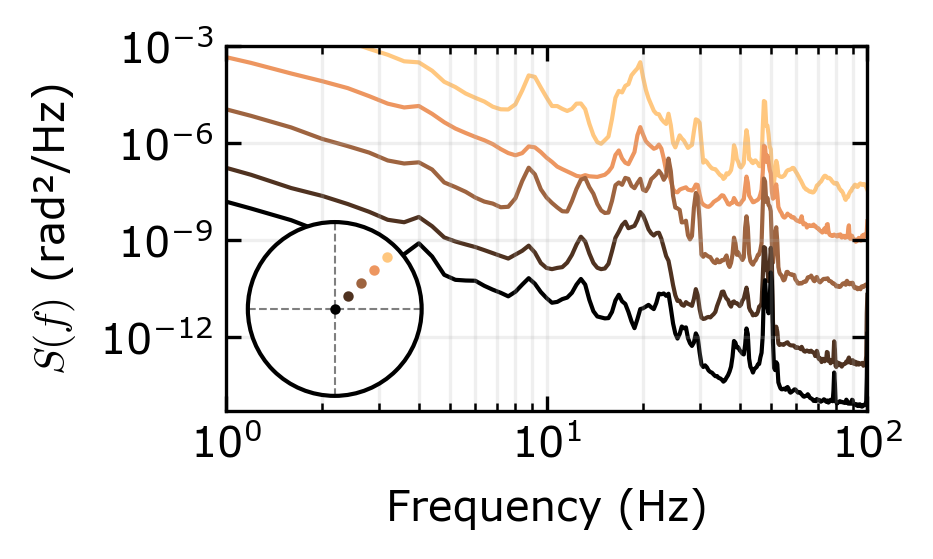}
    \caption{Variation of the inclination spectra with the position of the probe beam. For clarity, the curves are vertically offset by one order of magnitude each; the vertical scale applies to the lowest (black) trace. The beam's position within the $52$-mm diameter cell, measured with accuracy of 2~mm, is specified in the inset. In this series of experiments, $\mathrm{NA} = (12.5\pm0.3)\times10^{-3}$.}
    \label{fig:7}
\end{figure}

\begin{figure}[htbp]
    \centering
    \includegraphics[scale=1]{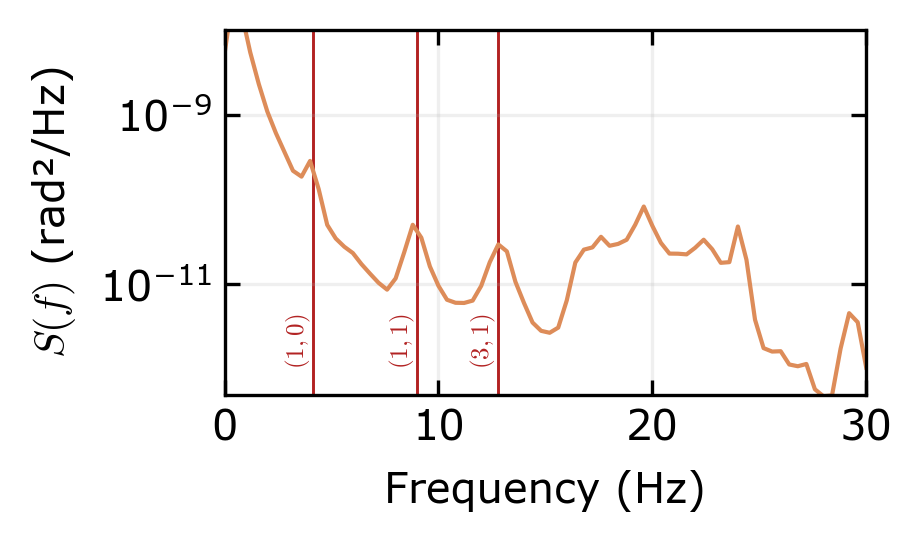}
    \caption{Surface inclination spectrum of water at $20^\circ$C, obtained by averaging the contributions from different positions of the incident beam relative to the sample cell shown in Fig.~\ref{fig:7}. Vertical red lines indicate the frequencies of surface eigenmodes expected in the cell satisfying the wetting boundary condition of Kidambi~\cite{kidambi2009capillary} that are closest to the three lowest peaks in the spectrum. Each line is labelled by its corresponding azimuthal and radial mode numbers $(m,n)$ of the wetting boundary condition.}
    \label{fig:8}
\end{figure}

Nonetheless, we explore these resonances further and look for their imprints across the sample cell. In Fig.~\ref{fig:7}, we record low-frequency fluctuations on the water surface for five different probe-beam positions, as indicated in the figure inset. Most spectral peaks are most pronounced when the probe beam is positioned near the cell edge (yellow line), suggesting that the waves are mechanically driven through vibrations of the cell walls. This behaviour is consistent with coupling between the liquid and the cell body mediated by the meniscus, where mechanical vibrations of the cell can generate waves from the edge by pulling on the liquid surface, i.e. meniscus waves. As the probe beam is moved gradually toward the centre of the cell, different peaks appear and disappear, reflecting the spatial structure of long-wavelength modes with varying eigenfunctions, as described by Eq.~\eqref{eq:psi}. Notably, when the probe beam is positioned exactly at the centre of the cell (black trace), only modes with azimuthal number $|m| = 1$ are expected to contribute to the surface inclination spectrum. This follows from the properties of Bessel functions at the origin, where $J^\prime_{m}(0) = 0$ except for $|m| = 1$. In practice, however, slight misalignment between the probe and the cell's geometric centre can lead to the appearance of additional peaks. For example, the peak near $12.80$~Hz, clearly resolved in the central spectrum shown in Fig.~\ref{fig:7}, corresponds to the $(m,n) = (3,1)$ mode (identified by frequency matching in Fig.~\ref{fig:8}), indicating the presence of such a residual offset in our data. The disappearance of $|m|\neq1$ modes thus represents an idealised case not fully realised in the present experiment, but it may offer a practical means of aligning the probe with the fluid sample centre.

The identification of \emph{specific} modes in the measured spectrum requires a prior knowledge of the realised boundary condition. However, accurate description of the fluid's behaviour at the boundary remains central to ongoing research in surface wave dynamics. Dussan first identified the so-called slip boundary condition \cite{dussan1976moving}, which arises when the contact line between the fluid and the container wall is free to move. This condition emerges above a critical wave amplitude, which can be experimentally determined by refracting a laser beam near the contact line \cite{coccairo1993experimental}. The presence of a slipping contact line alters both the eigenfrequency spectrum and the damping rate of surface waves \cite{kidambi2009meniscus}, which can be resolved in low-frequency SFSRS spectra as shifts in peak positions and changes in peak widths.

Subsequent studies revealed more intricate boundary behaviours, such as contact angle hysteresis and the sudden pinning of the fluid contact line \cite{viola2018capillary,bongarzone2024stick}. These combined effects of surface tension and viscosity were shown to further influence both the damping rate and frequency of surface waves \cite{viola2018theoretical}, requiring more precise readout of surface fluctuation spectra. Deviations from the expected dispersion relation \eqref{eq:disp} due to boundary conditions were recently confirmed experimentally in a waveguide configuration \cite{monsalve2022space-time}. In this context, SFSRS provides a powerful and accurate local probe of wave dynamics, capable of pinpointing wave frequencies and damping rates in the low-frequency regime. These results can be directly compared with other spatially-resolved techniques for the detection of low amplitude surface waves \cite{barroso2023digital,zhang2023pattern}.

We illustrate the mode identification process in Fig.~\ref{fig:8}, which shows the surface wave spectrum obtained by averaging signals from five beam positions (orange line). To associate the observed spectral peaks with the eigenfrequencies of the fluid sample, one must determine the corresponding wave numbers $k_{mn}$ by applying appropriate boundary conditions and evaluating the full dispersion relation given by Eq.~\eqref{eq:disp}. However, this task quickly becomes intractable, as the number of admissible modes increases rapidly with frequency. In the Supplemental document, we demonstrate this complexity for three different boundary conditions. To make the analysis manageable, we limit our focus to frequencies below $15$~Hz, where both the density of eigenmodes and viscous dissipation (which leads to peak broadening) are significantly reduced.

This approach enables us to associate the observed spectral peaks with specific surface modes predicted by the numerical model of Kidambi~\cite{kidambi2009capillary}, which incorporates a pinned contact-line boundary condition in a cylindrical geometry. We identify the three lowest-frequency peaks with the $(m,n) = (1,0)$ mode at $4.11$~Hz, the $(1,1)$ mode at $8.99$~Hz, and the $(3,1)$ mode at $12.80$~Hz, all indicated by red vertical lines in Fig.~\ref{fig:8}. We note that these represent the closest-matching eigenfrequencies; the remaining eight modes predicted in this frequency range appear to be unexcited in our system.

Establishing a firm correspondence between experimental data and a specific interface model requires the excitation of a broader range of surface modes. To obtain spectra with multiple low-frequency peaks suitable for comparison with theoretical predictions, these resonances must be more effectively excited, for instance by introducing specifically engineered source noise. Alternatively, reducing the size of the experimental cell generally increases the eigenfrequencies, thereby stretching the discrete wave spectrum over a broader frequency range. This strategy must be accompanied by improved frequency resolution of SFSRS, which could be achieved by extending the data acquisition time to enable denser sampling of the spectral peaks. On the other hand, viscous dissipation ultimately limits this approach. The peaks acquire a finite linewidth, bounded from below by the viscous dissipation rate $\gamma = 2(\mu/\rho)k^2$, which increases with wavenumber (frequency). The results shown in Fig.~\ref{fig:8} should therefore be regarded as a proof of principle, demonstrating a possible extension of SFSRS to low-frequency wave detection.

\section{Conclusions}
We have employed surface fluctuation specular reflection spectroscopy (SFSRS) to detect minute excitations of a liquid-air interface, covering wave frequencies across five orders of magnitude. Despite the reduced complexity of the experimental set-up, we successfully benchmarked the technique by detecting high-frequency capillary waves. The shape of the corresponding power spectral density of the interface inclination angle is consistent with the thermodynamic model of spontaneous thermal fluctuations, as previously reported \cite{aoki2018thermal,aoki2012spectral}. By precisely controlling the temperature of the liquid sample, we observed a clear temperature dependence of the thermal wave spectrum. However, distinguishing between the direct effects of temperature and those arising from the temperature dependence of the fluid's physical properties remains an open task.

At the low-frequency end of the spectrum, we identified discrete gravity-capillary wave modes emerging above the noise floor, excited stochastically by ambient vibrations. Spectral peaks corresponding to individual modes were resolved, although we argue that unambiguous identification is limited to the few lowest-frequency excitations, as higher modes quickly become too numerous to distinguish. We demonstrated that, in principle, low-frequency SFSRS can resolve these modes as isolated spectral features (cf. Fig.~\ref{fig:8}). The position and width of these peaks depend on the corresponding boundary condition, reflecting the behaviour of the fluid-wall contact line \cite{kidambi2009meniscus}. This highlights the potential of SFSRS as a valuable tool for probing boundary conditions in fluid systems.

Importantly, both low- and high-frequency wave regimes are accessible within the same experimental set-up, with the spectral bandwidth determined solely by the sampling rate and acquisition time. Because wave dynamics and excitation mechanisms depend strongly on the frequency range, detectors capable of spanning several orders of magnitude in frequency open new opportunities for exploring complex interfacial phenomena. For example, our results demonstrate that optical lever-based fluctuation sensors can probe the transition from a discrete "forest" of individual surface eigenmodes to a quasi-continuous excitation spectrum. The frequency at which this crossover occurs serves as a valuable probe of the underlying wave dynamics. As shown here, SFSRS offers a versatile, non-invasive approach for investigating such dynamics across a broad range of hydrodynamic and soft-matter systems.
\bigskip

\bmsection{Funding}
Engineering and Physical Sciences Research Council (EP/T517902/1); Science and Technology Facilities Council (ST/T006900/1);  Leverhulme Trust (RL-2019-020); Royal Society (UF120112).
\medskip

\bmsection{Acknowledgment}
We thank Chris Goodwin and Se\'{a}n Gregory from the Gravity Laboratory at the University of Nottingham for fruitful discussions and helpful suggestions. We are grateful to the technical staff of the School of Physics and Astronomy at the University of Nottingham for manufacturing the experimental cell central to this work. SCA is supported by the Engineering and Physical Sciences Research Council. P\v{S}, AJK and SW gratefully recognize funding from the Science and Technology Facilities Council through Quantum Simulators for Fundamental Physics, part of the UKRI Quantum Technologies for Fundamental Physics Program. VBS and SW gratefully acknowledge the support of the Leverhulme Research Leadership Award. SW also receives support from a Royal Society University Research Fellowship. Schemes of optical set-ups in this work were made with components obtained from the \emph{gwoptics Component Library} created by Alexander Franzen.
\medskip

\bmsection{Disclosures}
The authors declare no conflicts of interest.
\medskip

\bmsection{Data availability}
Data underlying the results presented in this paper are not publicly available at this time but may be obtained from the authors upon reasonable request.
\medskip

\bmsection{Supplemental document}
See Supplemental document for supporting content.

\interlinepenalty=10000
\bibliography{references}

\providecommand{\noopsort}[1]{}\providecommand{\singleletter}[1]{#1}%
\begin{thebibliography}{52}%
\makeatletter
\providecommand \@ifxundefined [1]{%
 \@ifx{#1\undefined}
}%
\providecommand \@ifnum [1]{%
 \ifnum #1\expandafter \@firstoftwo
 \else \expandafter \@secondoftwo
 \fi
}%
\providecommand \@ifx [1]{%
 \ifx #1\expandafter \@firstoftwo
 \else \expandafter \@secondoftwo
 \fi
}%
\providecommand \natexlab [1]{#1}%
\providecommand \enquote  [1]{``#1''}%
\providecommand \bibnamefont  [1]{#1}%
\providecommand \bibfnamefont [1]{#1}%
\providecommand \citenamefont [1]{#1}%
\providecommand \href@noop [0]{\@secondoftwo}%
\providecommand \href [0]{\begingroup \@sanitize@url \@href}%
\providecommand \@href[1]{\@@startlink{#1}\@@href}%
\providecommand \@@href[1]{\endgroup#1\@@endlink}%
\providecommand \@sanitize@url [0]{\catcode `\\12\catcode `\$12\catcode `\&12\catcode `\#12\catcode `\^12\catcode `\_12\catcode `\%12\relax}%
\providecommand \@@startlink[1]{}%
\providecommand \@@endlink[0]{}%
\providecommand \url  [0]{\begingroup\@sanitize@url \@url }%
\providecommand \@url [1]{\endgroup\@href {#1}{\urlprefix }}%
\providecommand \urlprefix  [0]{URL }%
\providecommand \Eprint [0]{\href }%
\providecommand \doibase [0]{https://doi.org/}%
\providecommand \selectlanguage [0]{\@gobble}%
\providecommand \bibinfo  [0]{\@secondoftwo}%
\providecommand \bibfield  [0]{\@secondoftwo}%
\providecommand \translation [1]{[#1]}%
\providecommand \BibitemOpen [0]{}%
\providecommand \bibitemStop [0]{}%
\providecommand \bibitemNoStop [0]{.\EOS\space}%
\providecommand \EOS [0]{\spacefactor3000\relax}%
\providecommand \BibitemShut  [1]{\csname bibitem#1\endcsname}%
\let\auto@bib@innerbib\@empty
\bibitem [{\citenamefont {Westlake}\ and\ \citenamefont {Cameron}(1967)}]{westlake1967fluid}%
  \BibitemOpen
  \bibfield  {author} {\bibinfo {author} {\bibfnamefont {F.}~\bibnamefont {Westlake}}\ and\ \bibinfo {author} {\bibfnamefont {A.}~\bibnamefont {Cameron}},\ }\bibfield  {title} {\bibinfo {title} {{Fluid film interferometry in lubrication studies}},\ }\href {https://doi.org/10.1038/214633a0} {\bibfield  {journal} {\bibinfo  {journal} {Nature}\ }\textbf {\bibinfo {volume} {214}},\ \bibinfo {pages} {633} (\bibinfo {year} {1967})}\BibitemShut {NoStop}%
\bibitem [{\citenamefont {Mitsui}\ and\ \citenamefont {Aoki}(2013)}]{mitsui2013measurements}%
  \BibitemOpen
  \bibfield  {author} {\bibinfo {author} {\bibfnamefont {T.}~\bibnamefont {Mitsui}}\ and\ \bibinfo {author} {\bibfnamefont {K.}~\bibnamefont {Aoki}},\ }\bibfield  {title} {\bibinfo {title} {Measurements of liquid surface fluctuations at sub-shot-noise levels with {Michelson} interferometry},\ }\href {https://doi.org/10.1103/PhysRevE.87.042403} {\bibfield  {journal} {\bibinfo  {journal} {Phys. Rev. E}\ }\textbf {\bibinfo {volume} {87}},\ \bibinfo {pages} {042403} (\bibinfo {year} {2013})}\BibitemShut {NoStop}%
\bibitem [{\citenamefont {Verma}\ and\ \citenamefont {Singh}(2015)}]{verma2015universal}%
  \BibitemOpen
  \bibfield  {author} {\bibinfo {author} {\bibfnamefont {G.}~\bibnamefont {Verma}}\ and\ \bibinfo {author} {\bibfnamefont {K.~P.}\ \bibnamefont {Singh}},\ }\bibfield  {title} {\bibinfo {title} {{Universal Long-Range Nanometric Bending of Water by Light}},\ }\href {https://doi.org/10.1103/PhysRevLett.115.143902} {\bibfield  {journal} {\bibinfo  {journal} {Phys. Rev. Lett.}\ }\textbf {\bibinfo {volume} {115}},\ \bibinfo {pages} {143902} (\bibinfo {year} {2015})}\BibitemShut {NoStop}%
\bibitem [{\citenamefont {Slavchov}\ \emph {et~al.}(2021)\citenamefont {Slavchov}, \citenamefont {Peychev},\ and\ \citenamefont {Ismail}}]{slavchov2021characterization}%
  \BibitemOpen
  \bibfield  {author} {\bibinfo {author} {\bibfnamefont {R.~I.}\ \bibnamefont {Slavchov}}, \bibinfo {author} {\bibfnamefont {B.}~\bibnamefont {Peychev}},\ and\ \bibinfo {author} {\bibfnamefont {A.~S.}\ \bibnamefont {Ismail}},\ }\bibfield  {title} {\bibinfo {title} {{Characterization of capillary waves: A review and a new optical method}},\ }\href {https://doi.org/10.1063/5.0066759} {\bibfield  {journal} {\bibinfo  {journal} {Phys. Fluids}\ }\textbf {\bibinfo {volume} {33}},\ \bibinfo {pages} {101303} (\bibinfo {year} {2021})}\BibitemShut {NoStop}%
\bibitem [{\citenamefont {Behroozi}\ \emph {et~al.}(2003)\citenamefont {Behroozi}, \citenamefont {Lambert},\ and\ \citenamefont {Buhrow}}]{behroozi2003noninvasive}%
  \BibitemOpen
  \bibfield  {author} {\bibinfo {author} {\bibfnamefont {F.}~\bibnamefont {Behroozi}}, \bibinfo {author} {\bibfnamefont {B.}~\bibnamefont {Lambert}},\ and\ \bibinfo {author} {\bibfnamefont {B.}~\bibnamefont {Buhrow}},\ }\bibfield  {title} {\bibinfo {title} {Noninvasive measurement of viscosity from damping of capillary waves},\ }\href {https://doi.org/10.1016/S0019-0578(07)60108-6} {\bibfield  {journal} {\bibinfo  {journal} {ISA Trans.}\ }\textbf {\bibinfo {volume} {42}},\ \bibinfo {pages} {3} (\bibinfo {year} {2003})}\BibitemShut {NoStop}%
\bibitem [{\citenamefont {Behroozi}\ \emph {et~al.}(2010)\citenamefont {Behroozi}, \citenamefont {Smith},\ and\ \citenamefont {Even}}]{behroozi2010stokes}%
  \BibitemOpen
  \bibfield  {author} {\bibinfo {author} {\bibfnamefont {F.}~\bibnamefont {Behroozi}}, \bibinfo {author} {\bibfnamefont {J.}~\bibnamefont {Smith}},\ and\ \bibinfo {author} {\bibfnamefont {W.}~\bibnamefont {Even}},\ }\bibfield  {title} {\bibinfo {title} {{Stokes’ dream: Measurement of fluid viscosity from the attenuation of capillary waves}},\ }\href {https://doi.org/10.1119/1.3467887} {\bibfield  {journal} {\bibinfo  {journal} {Am. J. Phys.}\ }\textbf {\bibinfo {volume} {78}},\ \bibinfo {pages} {1165} (\bibinfo {year} {2010})}\BibitemShut {NoStop}%
\bibitem [{\citenamefont {Tachizaki}\ \emph {et~al.}(2006)\citenamefont {Tachizaki}, \citenamefont {Muroya}, \citenamefont {Matsuda}, \citenamefont {Sugawara}, \citenamefont {Hurley},\ and\ \citenamefont {Wright}}]{tachizaki2006scanning}%
  \BibitemOpen
  \bibfield  {author} {\bibinfo {author} {\bibfnamefont {T.}~\bibnamefont {Tachizaki}}, \bibinfo {author} {\bibfnamefont {T.}~\bibnamefont {Muroya}}, \bibinfo {author} {\bibfnamefont {O.}~\bibnamefont {Matsuda}}, \bibinfo {author} {\bibfnamefont {Y.}~\bibnamefont {Sugawara}}, \bibinfo {author} {\bibfnamefont {D.~H.}\ \bibnamefont {Hurley}},\ and\ \bibinfo {author} {\bibfnamefont {O.~B.}\ \bibnamefont {Wright}},\ }\bibfield  {title} {\bibinfo {title} {Scanning ultrafast {Sagnac} interferometry for imaging two-dimensional surface wave propagation},\ }\href {https://doi.org/10.1063/1.2194518} {\bibfield  {journal} {\bibinfo  {journal} {Rev. Sci. Instrum.}\ }\textbf {\bibinfo {volume} {77}},\ \bibinfo {pages} {043713} (\bibinfo {year} {2006})}\BibitemShut {NoStop}%
\bibitem [{\citenamefont {Chaudhary}\ \emph {et~al.}(2021)\citenamefont {Chaudhary}, \citenamefont {Munjal},\ and\ \citenamefont {Singh}}]{chaudhary2021universal}%
  \BibitemOpen
  \bibfield  {author} {\bibinfo {author} {\bibfnamefont {K.}~\bibnamefont {Chaudhary}}, \bibinfo {author} {\bibfnamefont {P.}~\bibnamefont {Munjal}},\ and\ \bibinfo {author} {\bibfnamefont {K.~P.}\ \bibnamefont {Singh}},\ }\bibfield  {title} {\bibinfo {title} {{Universal {Stokes’s} nanomechanical viscometer}},\ }\href {https://doi.org/10.1038/s41598-021-93729-0} {\bibfield  {journal} {\bibinfo  {journal} {Sci. Rep.}\ }\textbf {\bibinfo {volume} {11}},\ \bibinfo {pages} {14365} (\bibinfo {year} {2021})}\BibitemShut {NoStop}%
\bibitem [{\citenamefont {Barroso}\ \emph {et~al.}(2023)\citenamefont {Barroso}, \citenamefont {Geelmuyden}, \citenamefont {Ajithkumar}, \citenamefont {Kent},\ and\ \citenamefont {Weinfurtner}}]{barroso2023digital}%
  \BibitemOpen
  \bibfield  {author} {\bibinfo {author} {\bibfnamefont {V.~S.}\ \bibnamefont {Barroso}}, \bibinfo {author} {\bibfnamefont {A.}~\bibnamefont {Geelmuyden}}, \bibinfo {author} {\bibfnamefont {S.~C.}\ \bibnamefont {Ajithkumar}}, \bibinfo {author} {\bibfnamefont {A.~J.}\ \bibnamefont {Kent}},\ and\ \bibinfo {author} {\bibfnamefont {S.}~\bibnamefont {Weinfurtner}},\ }\bibfield  {title} {\bibinfo {title} {Multiplexed digital holography for fluid surface profilometry},\ }\href {https://doi.org/10.1364/AO.496937} {\bibfield  {journal} {\bibinfo  {journal} {Appl. Opt.}\ }\textbf {\bibinfo {volume} {62}},\ \bibinfo {pages} {7175} (\bibinfo {year} {2023})}\BibitemShut {NoStop}%
\bibitem [{\citenamefont {Shmyrov}\ \emph {et~al.}(2019)\citenamefont {Shmyrov}, \citenamefont {Mizev}, \citenamefont {Shmyrova},\ and\ \citenamefont {Mizeva}}]{shmyrov2019capillary}%
  \BibitemOpen
  \bibfield  {author} {\bibinfo {author} {\bibfnamefont {A.}~\bibnamefont {Shmyrov}}, \bibinfo {author} {\bibfnamefont {A.}~\bibnamefont {Mizev}}, \bibinfo {author} {\bibfnamefont {A.}~\bibnamefont {Shmyrova}},\ and\ \bibinfo {author} {\bibfnamefont {I.}~\bibnamefont {Mizeva}},\ }\bibfield  {title} {\bibinfo {title} {{Capillary wave method: An alternative approach to wave excitation and to wave profile reconstruction}},\ }\href {https://doi.org/10.1063/1.5060666} {\bibfield  {journal} {\bibinfo  {journal} {Phys. Fluids}\ }\textbf {\bibinfo {volume} {31}},\ \bibinfo {pages} {012101} (\bibinfo {year} {2019})}\BibitemShut {NoStop}%
\bibitem [{\citenamefont {Katyl}\ and\ \citenamefont {Ingard}(1967)}]{katyl1967line}%
  \BibitemOpen
  \bibfield  {author} {\bibinfo {author} {\bibfnamefont {R.~H.}\ \bibnamefont {Katyl}}\ and\ \bibinfo {author} {\bibfnamefont {U.}~\bibnamefont {Ingard}},\ }\bibfield  {title} {\bibinfo {title} {Line broadening of light scattered from a liquid surface},\ }\href {https://doi.org/10.1103/PhysRevLett.19.64} {\bibfield  {journal} {\bibinfo  {journal} {Phys. Rev. Lett.}\ }\textbf {\bibinfo {volume} {19}},\ \bibinfo {pages} {64} (\bibinfo {year} {1967})}\BibitemShut {NoStop}%
\bibitem [{\citenamefont {Katyl}\ and\ \citenamefont {Ingard}(1968)}]{katyl1968scattering}%
  \BibitemOpen
  \bibfield  {author} {\bibinfo {author} {\bibfnamefont {R.~H.}\ \bibnamefont {Katyl}}\ and\ \bibinfo {author} {\bibfnamefont {U.}~\bibnamefont {Ingard}},\ }\bibfield  {title} {\bibinfo {title} {Scattering of light by thermal ripplons},\ }\href {https://doi.org/10.1103/PhysRevLett.20.248} {\bibfield  {journal} {\bibinfo  {journal} {Phys. Rev. Lett.}\ }\textbf {\bibinfo {volume} {20}},\ \bibinfo {pages} {248} (\bibinfo {year} {1968})}\BibitemShut {NoStop}%
\bibitem [{\citenamefont {Bouchiat}\ \emph {et~al.}(1968)\citenamefont {Bouchiat}, \citenamefont {Meunier},\ and\ \citenamefont {Brossel}}]{bouchiat1968demonstration}%
  \BibitemOpen
  \bibfield  {author} {\bibinfo {author} {\bibfnamefont {M.}~\bibnamefont {Bouchiat}}, \bibinfo {author} {\bibfnamefont {J.}~\bibnamefont {Meunier}},\ and\ \bibinfo {author} {\bibfnamefont {J.}~\bibnamefont {Brossel}},\ }\bibfield  {title} {\bibinfo {title} {Demonstration of capillarity waves excited thermically at surface of a liquid by inelastic diffusion of coherent light},\ }\href@noop {} {\bibfield  {journal} {\bibinfo  {journal} {C. R. Acad. Sci. B}\ }\textbf {\bibinfo {volume} {266}},\ \bibinfo {pages} {255} (\bibinfo {year} {1968})}\BibitemShut {NoStop}%
\bibitem [{\citenamefont {Kubo}(1966)}]{kubo1966fluctuation}%
  \BibitemOpen
  \bibfield  {author} {\bibinfo {author} {\bibfnamefont {R.}~\bibnamefont {Kubo}},\ }\bibfield  {title} {\bibinfo {title} {The fluctuation-dissipation theorem},\ }\href {https://doi.org/10.1088/0034-4885/29/1/306} {\bibfield  {journal} {\bibinfo  {journal} {Rep. Prog. Phys.}\ }\textbf {\bibinfo {volume} {29}},\ \bibinfo {pages} {255} (\bibinfo {year} {1966})}\BibitemShut {NoStop}%
\bibitem [{\citenamefont {Huang}\ and\ \citenamefont {Webb}(1969)}]{huang1969viscous}%
  \BibitemOpen
  \bibfield  {author} {\bibinfo {author} {\bibfnamefont {J.}~\bibnamefont {Huang}}\ and\ \bibinfo {author} {\bibfnamefont {W.}~\bibnamefont {Webb}},\ }\bibfield  {title} {\bibinfo {title} {Viscous damping of thermal excitations on the interface of critical fluid mixtures},\ }\href {https://doi.org/10.1103/PhysRevLett.23.160} {\bibfield  {journal} {\bibinfo  {journal} {Phys. Rev. Lett.}\ }\textbf {\bibinfo {volume} {23}},\ \bibinfo {pages} {160} (\bibinfo {year} {1969})}\BibitemShut {NoStop}%
\bibitem [{\citenamefont {Meunier}(1969)}]{meunier1969diffusion}%
  \BibitemOpen
  \bibfield  {author} {\bibinfo {author} {\bibfnamefont {J.}~\bibnamefont {Meunier}},\ }\bibfield  {title} {\bibinfo {title} {Diffusion de la lumi{\`e}re par les ondes de surface sur {CO\textsubscript{2}} pr{\`e}s du point critique mesure de la tension superficielle},\ }\href@noop {} {\bibfield  {journal} {\bibinfo  {journal} {J. Phys.}\ }\textbf {\bibinfo {volume} {30}},\ \bibinfo {pages} {933} (\bibinfo {year} {1969})}\BibitemShut {NoStop}%
\bibitem [{\citenamefont {Sohl}\ \emph {et~al.}(1978)\citenamefont {Sohl}, \citenamefont {Miyano},\ and\ \citenamefont {Ketterson}}]{sohl1978novel}%
  \BibitemOpen
  \bibfield  {author} {\bibinfo {author} {\bibfnamefont {C.~H.}\ \bibnamefont {Sohl}}, \bibinfo {author} {\bibfnamefont {K.}~\bibnamefont {Miyano}},\ and\ \bibinfo {author} {\bibfnamefont {J.~B.}\ \bibnamefont {Ketterson}},\ }\bibfield  {title} {\bibinfo {title} {Novel technique for dynamic surface tension and viscosity measurements at liquid–gas interfaces},\ }\href {https://doi.org/10.1063/1.1135288} {\bibfield  {journal} {\bibinfo  {journal} {Rev. Sci. Instrum.}\ }\textbf {\bibinfo {volume} {49}},\ \bibinfo {pages} {1464} (\bibinfo {year} {1978})}\BibitemShut {NoStop}%
\bibitem [{\citenamefont {Huang}\ and\ \citenamefont {Wang}(1998)}]{huang1998effects}%
  \BibitemOpen
  \bibfield  {author} {\bibinfo {author} {\bibfnamefont {Q.~R.}\ \bibnamefont {Huang}}\ and\ \bibinfo {author} {\bibfnamefont {C.~H.}\ \bibnamefont {Wang}},\ }\bibfield  {title} {\bibinfo {title} {Effects of viscoelasticity of bulk polymer solution on the surface modes as probed by laser light scattering},\ }\href {https://doi.org/10.1063/1.477237} {\bibfield  {journal} {\bibinfo  {journal} {J. Chem. Phys.}\ }\textbf {\bibinfo {volume} {109}},\ \bibinfo {pages} {6103} (\bibinfo {year} {1998})}\BibitemShut {NoStop}%
\bibitem [{\citenamefont {Cicuta}\ and\ \citenamefont {Hopkinson}(2004)}]{cicuta2004recent}%
  \BibitemOpen
  \bibfield  {author} {\bibinfo {author} {\bibfnamefont {P.}~\bibnamefont {Cicuta}}\ and\ \bibinfo {author} {\bibfnamefont {I.}~\bibnamefont {Hopkinson}},\ }\bibfield  {title} {\bibinfo {title} {Recent developments of surface light scattering as a tool for optical-rheology of polymer monolayers},\ }\href {https://doi.org/10.1016/j.colsurfa.2003.11.025} {\bibfield  {journal} {\bibinfo  {journal} {Colloids Surf. A}\ }\textbf {\bibinfo {volume} {233}},\ \bibinfo {pages} {97} (\bibinfo {year} {2004})}\BibitemShut {NoStop}%
\bibitem [{\citenamefont {Streit-Nierobisch}\ \emph {et~al.}(2008)\citenamefont {Streit-Nierobisch}, \citenamefont {Gutt}, \citenamefont {Paulus},\ and\ \citenamefont {Tolan}}]{streit2008cooling}%
  \BibitemOpen
  \bibfield  {author} {\bibinfo {author} {\bibfnamefont {S.}~\bibnamefont {Streit-Nierobisch}}, \bibinfo {author} {\bibfnamefont {C.}~\bibnamefont {Gutt}}, \bibinfo {author} {\bibfnamefont {M.}~\bibnamefont {Paulus}},\ and\ \bibinfo {author} {\bibfnamefont {M.}~\bibnamefont {Tolan}},\ }\bibfield  {title} {\bibinfo {title} {Cooling rate dependence of the glass transition at free surfaces},\ }\href {https://doi.org/10.1103/PhysRevB.77.041410} {\bibfield  {journal} {\bibinfo  {journal} {Phys. Rev. B}\ }\textbf {\bibinfo {volume} {77}},\ \bibinfo {pages} {041410} (\bibinfo {year} {2008})}\BibitemShut {NoStop}%
\bibitem [{\citenamefont {Kikuchi}\ \emph {et~al.}(1994)\citenamefont {Kikuchi}, \citenamefont {Sakai},\ and\ \citenamefont {Takagi}}]{kikuchi1994complex}%
  \BibitemOpen
  \bibfield  {author} {\bibinfo {author} {\bibfnamefont {H.}~\bibnamefont {Kikuchi}}, \bibinfo {author} {\bibfnamefont {K.}~\bibnamefont {Sakai}},\ and\ \bibinfo {author} {\bibfnamefont {K.}~\bibnamefont {Takagi}},\ }\bibfield  {title} {\bibinfo {title} {Complex propagation of surface waves on soft gels},\ }\href {https://doi.org/10.1103/PhysRevB.49.3061} {\bibfield  {journal} {\bibinfo  {journal} {Phys. Rev. B}\ }\textbf {\bibinfo {volume} {49}},\ \bibinfo {pages} {3061} (\bibinfo {year} {1994})}\BibitemShut {NoStop}%
\bibitem [{\citenamefont {Ahn}\ \emph {et~al.}(2001)\citenamefont {Ahn}, \citenamefont {Yoon},\ and\ \citenamefont {Kim}}]{keunho2001thickness}%
  \BibitemOpen
  \bibfield  {author} {\bibinfo {author} {\bibfnamefont {K.}~\bibnamefont {Ahn}}, \bibinfo {author} {\bibfnamefont {K.~H.}\ \bibnamefont {Yoon}},\ and\ \bibinfo {author} {\bibfnamefont {M.~W.}\ \bibnamefont {Kim}},\ }\bibfield  {title} {\bibinfo {title} {Thickness dependence of surface modes on a gel},\ }\href {https://doi.org/10.1209/epl/i2001-00295-7} {\bibfield  {journal} {\bibinfo  {journal} {EPL}\ }\textbf {\bibinfo {volume} {54}},\ \bibinfo {pages} {199} (\bibinfo {year} {2001})}\BibitemShut {NoStop}%
\bibitem [{\citenamefont {Langevin}(2021)}]{langevin2021light}%
  \BibitemOpen
  \bibfield  {author} {\bibinfo {author} {\bibfnamefont {D.}~\bibnamefont {Langevin}},\ }\bibfield  {title} {\bibinfo {title} {Light scattering by liquid surfaces, new developments},\ }\href {https://doi.org/10.1016/j.cis.2021.102368} {\bibfield  {journal} {\bibinfo  {journal} {Adv. Colloid Interface Sci.}\ }\textbf {\bibinfo {volume} {289}},\ \bibinfo {pages} {102368} (\bibinfo {year} {2021})}\BibitemShut {NoStop}%
\bibitem [{\citenamefont {Tay}\ \emph {et~al.}(2008)\citenamefont {Tay}, \citenamefont {Thibierge}, \citenamefont {Fournier}, \citenamefont {Fretigny}, \citenamefont {Lequeux}, \citenamefont {Monteux}, \citenamefont {Roger},\ and\ \citenamefont {Talini}}]{tay2008probing}%
  \BibitemOpen
  \bibfield  {author} {\bibinfo {author} {\bibfnamefont {A.}~\bibnamefont {Tay}}, \bibinfo {author} {\bibfnamefont {C.}~\bibnamefont {Thibierge}}, \bibinfo {author} {\bibfnamefont {D.}~\bibnamefont {Fournier}}, \bibinfo {author} {\bibfnamefont {C.}~\bibnamefont {Fretigny}}, \bibinfo {author} {\bibfnamefont {F.}~\bibnamefont {Lequeux}}, \bibinfo {author} {\bibfnamefont {C.}~\bibnamefont {Monteux}}, \bibinfo {author} {\bibfnamefont {J.~P.}\ \bibnamefont {Roger}},\ and\ \bibinfo {author} {\bibfnamefont {L.}~\bibnamefont {Talini}},\ }\bibfield  {title} {\bibinfo {title} {Probing thermal waves on the free surface of various media: Surface fluctuation specular reflection spectroscopy},\ }\href {https://doi.org/10.1063/1.3002424} {\bibfield  {journal} {\bibinfo  {journal} {Rev. Sci. Instrum.}\ }\textbf {\bibinfo {volume} {79}},\ \bibinfo {pages} {103107} (\bibinfo {year} {2008})}\BibitemShut {NoStop}%
\bibitem [{\citenamefont {Pottier}\ \emph {et~al.}(2011)\citenamefont {Pottier}, \citenamefont {Ducouret}, \citenamefont {Fr{\'e}tigny}, \citenamefont {Lequeux},\ and\ \citenamefont {Talini}}]{pottier2011high}%
  \BibitemOpen
  \bibfield  {author} {\bibinfo {author} {\bibfnamefont {B.}~\bibnamefont {Pottier}}, \bibinfo {author} {\bibfnamefont {G.}~\bibnamefont {Ducouret}}, \bibinfo {author} {\bibfnamefont {C.}~\bibnamefont {Fr{\'e}tigny}}, \bibinfo {author} {\bibfnamefont {F.}~\bibnamefont {Lequeux}},\ and\ \bibinfo {author} {\bibfnamefont {L.}~\bibnamefont {Talini}},\ }\bibfield  {title} {\bibinfo {title} {High bandwidth linear viscoelastic properties of complex fluids from the measurement of their free surface fluctuations},\ }\href {https://doi.org/10.1039/C1SM05258F} {\bibfield  {journal} {\bibinfo  {journal} {Soft Matter}\ }\textbf {\bibinfo {volume} {7}},\ \bibinfo {pages} {7843} (\bibinfo {year} {2011})}\BibitemShut {NoStop}%
\bibitem [{\citenamefont {Aoki}\ and\ \citenamefont {Mitsui}(2018)}]{aoki2018thermal}%
  \BibitemOpen
  \bibfield  {author} {\bibinfo {author} {\bibfnamefont {K.}~\bibnamefont {Aoki}}\ and\ \bibinfo {author} {\bibfnamefont {T.}~\bibnamefont {Mitsui}},\ }\bibfield  {title} {\bibinfo {title} {Thermal interface fluctuations of liquids and viscoelastic materials},\ }\href {https://doi.org/10.1093/ptep/pty026} {\bibfield  {journal} {\bibinfo  {journal} {Prog. Theor. Exp. Phys.}\ }\textbf {\bibinfo {volume} {2018}},\ \bibinfo {pages} {043J01} (\bibinfo {year} {2018})}\BibitemShut {NoStop}%
\bibitem [{\citenamefont {Pottier}\ \emph {et~al.}(2015)\citenamefont {Pottier}, \citenamefont {Fr{\'e}tigny},\ and\ \citenamefont {Talini}}]{pottier2015boundary}%
  \BibitemOpen
  \bibfield  {author} {\bibinfo {author} {\bibfnamefont {B.}~\bibnamefont {Pottier}}, \bibinfo {author} {\bibfnamefont {C.}~\bibnamefont {Fr{\'e}tigny}},\ and\ \bibinfo {author} {\bibfnamefont {L.}~\bibnamefont {Talini}},\ }\bibfield  {title} {\bibinfo {title} {Boundary condition in liquid thin films revealed through the thermal fluctuations of their free surfaces},\ }\href {https://doi.org/10.1103/PhysRevLett.114.227801} {\bibfield  {journal} {\bibinfo  {journal} {Phys. Rev. Lett.}\ }\textbf {\bibinfo {volume} {114}},\ \bibinfo {pages} {227801} (\bibinfo {year} {2015})}\BibitemShut {NoStop}%
\bibitem [{\citenamefont {Mitsui}\ and\ \citenamefont {Aoki}(2019)}]{mitsui2019fluctuation}%
  \BibitemOpen
  \bibfield  {author} {\bibinfo {author} {\bibfnamefont {T.}~\bibnamefont {Mitsui}}\ and\ \bibinfo {author} {\bibfnamefont {K.}~\bibnamefont {Aoki}},\ }\bibfield  {title} {\bibinfo {title} {Fluctuation spectroscopy of surface melting of ice with and without impurities},\ }\href {https://doi.org/10.1103/PhysRevE.99.010801} {\bibfield  {journal} {\bibinfo  {journal} {Phys. Rev. E}\ }\textbf {\bibinfo {volume} {99}},\ \bibinfo {pages} {010801} (\bibinfo {year} {2019})}\BibitemShut {NoStop}%
\bibitem [{\citenamefont {Raudsepp}\ \emph {et~al.}(2012)\citenamefont {Raudsepp}, \citenamefont {Fretigny}, \citenamefont {Lequeux},\ and\ \citenamefont {Talini}}]{raudsepp2012two}%
  \BibitemOpen
  \bibfield  {author} {\bibinfo {author} {\bibfnamefont {A.}~\bibnamefont {Raudsepp}}, \bibinfo {author} {\bibfnamefont {C.}~\bibnamefont {Fretigny}}, \bibinfo {author} {\bibfnamefont {F.}~\bibnamefont {Lequeux}},\ and\ \bibinfo {author} {\bibfnamefont {L.}~\bibnamefont {Talini}},\ }\bibfield  {title} {\bibinfo {title} {Two beam surface fluctuation specular reflection spectroscopy},\ }\href {https://doi.org/10.1063/1.3678317} {\bibfield  {journal} {\bibinfo  {journal} {Rev. Sci. Instrum.}\ }\textbf {\bibinfo {volume} {83}},\ \bibinfo {pages} {013111} (\bibinfo {year} {2012})}\BibitemShut {NoStop}%
\bibitem [{\citenamefont {Hobbs}(1997)}]{hobbs1997ultrasensitive}%
  \BibitemOpen
  \bibfield  {author} {\bibinfo {author} {\bibfnamefont {P.~C.~D.}\ \bibnamefont {Hobbs}},\ }\bibfield  {title} {\bibinfo {title} {Ultrasensitive laser measurements without tears},\ }\href {https://doi.org/10.1364/AO.36.000903} {\bibfield  {journal} {\bibinfo  {journal} {Applied Optics}\ }\textbf {\bibinfo {volume} {36}},\ \bibinfo {pages} {903–920} (\bibinfo {year} {1997})}\BibitemShut {NoStop}%
\bibitem [{\citenamefont {Ajithkumar}(2025)}]{sree2025thesis}%
  \BibitemOpen
  \bibfield  {author} {\bibinfo {author} {\bibfnamefont {S.~C.}\ \bibnamefont {Ajithkumar}},\ }\emph {\bibinfo {title} {Optical Sensing of Classical and Quantum Fluid Interfaces}},\ \href@noop {} {Ph.D. thesis},\ \bibinfo  {school} {University of Nottingham} (\bibinfo {year} {2025})\BibitemShut {NoStop}%
\bibitem [{\citenamefont {Henderson}\ and\ \citenamefont {Miles}(1994)}]{Henderson1994Surface-waveLine}%
  \BibitemOpen
  \bibfield  {author} {\bibinfo {author} {\bibfnamefont {D.~M.}\ \bibnamefont {Henderson}}\ and\ \bibinfo {author} {\bibfnamefont {J.~W.}\ \bibnamefont {Miles}},\ }\bibfield  {title} {\bibinfo {title} {{Surface-wave damping in a circular cylinder with a fixed contact line}},\ }\href {https://doi.org/10.1017/S0022112094002363} {\bibfield  {journal} {\bibinfo  {journal} {J. Fluid Mech.}\ }\textbf {\bibinfo {volume} {275}},\ \bibinfo {pages} {285} (\bibinfo {year} {1994})}\BibitemShut {NoStop}%
\bibitem [{\citenamefont {Coccairo}\ \emph {et~al.}(1993)\citenamefont {Coccairo}, \citenamefont {Faetti},\ and\ \citenamefont {Festa}}]{coccairo1993experimental}%
  \BibitemOpen
  \bibfield  {author} {\bibinfo {author} {\bibfnamefont {B.}~\bibnamefont {Coccairo}}, \bibinfo {author} {\bibfnamefont {S.}~\bibnamefont {Faetti}},\ and\ \bibinfo {author} {\bibfnamefont {C.}~\bibnamefont {Festa}},\ }\bibfield  {title} {\bibinfo {title} {Experimental investigation of capillarity effects on surface gravity waves: non-wetting boundary conditions},\ }\href {https://doi.org/10.1017/S0022112093000035} {\bibfield  {journal} {\bibinfo  {journal} {J. Fluid Mech.}\ }\textbf {\bibinfo {volume} {246}},\ \bibinfo {pages} {43} (\bibinfo {year} {1993})}\BibitemShut {NoStop}%
\bibitem [{\citenamefont {Picard}\ and\ \citenamefont {Davoust}(2007)}]{picard2007resonance}%
  \BibitemOpen
  \bibfield  {author} {\bibinfo {author} {\bibfnamefont {C.}~\bibnamefont {Picard}}\ and\ \bibinfo {author} {\bibfnamefont {L.}~\bibnamefont {Davoust}},\ }\bibfield  {title} {\bibinfo {title} {{Resonance Frequencies of Meniscus Waves as a Physical Mechanism for a {DNA} Biosensor}},\ }\href {https://doi.org/10.1021/la0624236} {\bibfield  {journal} {\bibinfo  {journal} {Langmuir}\ }\textbf {\bibinfo {volume} {23}},\ \bibinfo {pages} {1394} (\bibinfo {year} {2007})}\BibitemShut {NoStop}%
\bibitem [{\citenamefont {Howell}\ \emph {et~al.}(2000)\citenamefont {Howell}, \citenamefont {Buhrow}, \citenamefont {Heath}, \citenamefont {McKenna}, \citenamefont {Hwang},\ and\ \citenamefont {Schatz}}]{howell2000measurements}%
  \BibitemOpen
  \bibfield  {author} {\bibinfo {author} {\bibfnamefont {D.~R.}\ \bibnamefont {Howell}}, \bibinfo {author} {\bibfnamefont {B.}~\bibnamefont {Buhrow}}, \bibinfo {author} {\bibfnamefont {T.}~\bibnamefont {Heath}}, \bibinfo {author} {\bibfnamefont {C.}~\bibnamefont {McKenna}}, \bibinfo {author} {\bibfnamefont {W.}~\bibnamefont {Hwang}},\ and\ \bibinfo {author} {\bibfnamefont {M.~F.}\ \bibnamefont {Schatz}},\ }\bibfield  {title} {\bibinfo {title} {{Measurements of surface-wave damping in a container}},\ }\href {https://doi.org/10.1063/1.870310} {\bibfield  {journal} {\bibinfo  {journal} {Phys. Fluids}\ }\textbf {\bibinfo {volume} {12}},\ \bibinfo {pages} {322} (\bibinfo {year} {2000})}\BibitemShut {NoStop}%
\bibitem [{\citenamefont {Kidambi}(2009{\natexlab{a}})}]{kidambi2009capillary}%
  \BibitemOpen
  \bibfield  {author} {\bibinfo {author} {\bibfnamefont {R.}~\bibnamefont {Kidambi}},\ }\bibfield  {title} {\bibinfo {title} {{Capillary damping of inviscid surface waves in a circular cylinder}},\ }\href {https://doi.org/DOI: 10.1017/S0022112009005898} {\bibfield  {journal} {\bibinfo  {journal} {J. Fluid Mech.}\ }\textbf {\bibinfo {volume} {627}},\ \bibinfo {pages} {323} (\bibinfo {year} {2009}{\natexlab{a}})}\BibitemShut {NoStop}%
\bibitem [{\citenamefont {Kidambi}(2009{\natexlab{b}})}]{kidambi2009meniscus}%
  \BibitemOpen
  \bibfield  {author} {\bibinfo {author} {\bibfnamefont {R.}~\bibnamefont {Kidambi}},\ }\bibfield  {title} {\bibinfo {title} {Meniscus effects on the frequency and damping of capillary-gravity waves in a brimful circular cylinder},\ }\href {https://doi.org/10.1016/j.wavemoti.2008.10.001} {\bibfield  {journal} {\bibinfo  {journal} {Wave Motion}\ }\textbf {\bibinfo {volume} {46}},\ \bibinfo {pages} {144–154} (\bibinfo {year} {2009}{\natexlab{b}})}\BibitemShut {NoStop}%
\bibitem [{\citenamefont {Aoki}\ and\ \citenamefont {Mitsui}(2012)}]{aoki2012spectral}%
  \BibitemOpen
  \bibfield  {author} {\bibinfo {author} {\bibfnamefont {K.}~\bibnamefont {Aoki}}\ and\ \bibinfo {author} {\bibfnamefont {T.}~\bibnamefont {Mitsui}},\ }\bibfield  {title} {\bibinfo {title} {{Spectral properties of thermal fluctuations on simple liquid surfaces below shot-noise levels}},\ }\href {https://doi.org/10.1103/PhysRevE.86.011602} {\bibfield  {journal} {\bibinfo  {journal} {Phys. Rev. E}\ }\textbf {\bibinfo {volume} {86}},\ \bibinfo {pages} {011602} (\bibinfo {year} {2012})}\BibitemShut {NoStop}%
\bibitem [{\citenamefont {Bouchiat}\ and\ \citenamefont {Meunier}(1971)}]{bouchiat1971spectre}%
  \BibitemOpen
  \bibfield  {author} {\bibinfo {author} {\bibfnamefont {M.~A.}\ \bibnamefont {Bouchiat}}\ and\ \bibinfo {author} {\bibfnamefont {J.}~\bibnamefont {Meunier}},\ }\bibfield  {title} {\bibinfo {title} {Spectre des fluctuations thermiques de la surface libre d’un liquide simple},\ }\href {https://doi.org/10.1051/jphys:01971003207056100} {\bibfield  {journal} {\bibinfo  {journal} {J. Phys.}\ }\textbf {\bibinfo {volume} {32}},\ \bibinfo {pages} {561–571} (\bibinfo {year} {1971})}\BibitemShut {NoStop}%
\bibitem [{\citenamefont {Lemmon}\ \emph {et~al.}(2025)\citenamefont {Lemmon}, \citenamefont {Bell}, \citenamefont {Huber},\ and\ \citenamefont {McLinden}}]{water-data}%
  \BibitemOpen
  \bibfield  {author} {\bibinfo {author} {\bibfnamefont {E.~W.}\ \bibnamefont {Lemmon}}, \bibinfo {author} {\bibfnamefont {I.~H.}\ \bibnamefont {Bell}}, \bibinfo {author} {\bibfnamefont {M.~L.}\ \bibnamefont {Huber}},\ and\ \bibinfo {author} {\bibfnamefont {M.~O.}\ \bibnamefont {McLinden}},\ }\bibinfo {title} {Nist chemistry webbook, nist standard reference database number 69}\ (\bibinfo  {publisher} {National Institute of Standards and Technology},\ \bibinfo {year} {2025})\ Chap.\ \bibinfo {chapter} {Thermophysical Properties of Fluid Systems}\BibitemShut {NoStop}%
\bibitem [{\citenamefont {Gon\c{c}alves}\ \emph {et~al.}(2010)\citenamefont {Gon\c{c}alves}, \citenamefont {Trindade}, \citenamefont {Costa}, \citenamefont {Bernardo}, \citenamefont {Johnson}, \citenamefont {Fonseca},\ and\ \citenamefont {Ferreira}}]{ethanol-data}%
  \BibitemOpen
  \bibfield  {author} {\bibinfo {author} {\bibfnamefont {F.}~\bibnamefont {Gon\c{c}alves}}, \bibinfo {author} {\bibfnamefont {A.}~\bibnamefont {Trindade}}, \bibinfo {author} {\bibfnamefont {C.}~\bibnamefont {Costa}}, \bibinfo {author} {\bibfnamefont {J.}~\bibnamefont {Bernardo}}, \bibinfo {author} {\bibfnamefont {I.}~\bibnamefont {Johnson}}, \bibinfo {author} {\bibfnamefont {I.}~\bibnamefont {Fonseca}},\ and\ \bibinfo {author} {\bibfnamefont {A.}~\bibnamefont {Ferreira}},\ }\bibfield  {title} {\bibinfo {title} {{PVT}, viscosity, and surface tension of ethanol: {New} measurements and literature data evaluation},\ }\href {https://doi.org/10.1016/j.jct.2010.03.022} {\bibfield  {journal} {\bibinfo  {journal} {J. Chem. Thermodyn.}\ }\textbf {\bibinfo {volume} {42}},\ \bibinfo {pages} {1039–1049} (\bibinfo {year} {2010})}\BibitemShut {NoStop}%
\bibitem [{\citenamefont {Takamura}\ \emph {et~al.}(2012)\citenamefont {Takamura}, \citenamefont {Fischer},\ and\ \citenamefont {Morrow}}]{Takamura2012}%
  \BibitemOpen
  \bibfield  {author} {\bibinfo {author} {\bibfnamefont {K.}~\bibnamefont {Takamura}}, \bibinfo {author} {\bibfnamefont {H.}~\bibnamefont {Fischer}},\ and\ \bibinfo {author} {\bibfnamefont {N.~R.}\ \bibnamefont {Morrow}},\ }\bibfield  {title} {\bibinfo {title} {Physical properties of aqueous glycerol solutions},\ }\href {https://doi.org/10.1016/j.petrol.2012.09.003} {\bibfield  {journal} {\bibinfo  {journal} {J. Pet. Eng.}\ }\textbf {\bibinfo {volume} {98–99}},\ \bibinfo {pages} {50–60} (\bibinfo {year} {2012})}\BibitemShut {NoStop}%
\bibitem [{\citenamefont {Dussan~V.}(1976)}]{dussan1976moving}%
  \BibitemOpen
  \bibfield  {author} {\bibinfo {author} {\bibfnamefont {E.~B.}\ \bibnamefont {Dussan~V.}},\ }\bibfield  {title} {\bibinfo {title} {The moving contact line: the slip boundary condition},\ }\href {https://doi.org/10.1017/S0022112076002838} {\bibfield  {journal} {\bibinfo  {journal} {J. Fluid Mech.}\ }\textbf {\bibinfo {volume} {77}},\ \bibinfo {pages} {665} (\bibinfo {year} {1976})}\BibitemShut {NoStop}%
\bibitem [{\citenamefont {Viola}\ \emph {et~al.}(2018)\citenamefont {Viola}, \citenamefont {Brun},\ and\ \citenamefont {Gallaire}}]{viola2018capillary}%
  \BibitemOpen
  \bibfield  {author} {\bibinfo {author} {\bibfnamefont {F.}~\bibnamefont {Viola}}, \bibinfo {author} {\bibfnamefont {P.-T.}\ \bibnamefont {Brun}},\ and\ \bibinfo {author} {\bibfnamefont {F.}~\bibnamefont {Gallaire}},\ }\bibfield  {title} {\bibinfo {title} {Capillary hysteresis in sloshing dynamics: a weakly nonlinear analysis},\ }\href {https://doi.org/10.1017/jfm.2017.860} {\bibfield  {journal} {\bibinfo  {journal} {J. Fluid Mech.}\ }\textbf {\bibinfo {volume} {837}},\ \bibinfo {pages} {788} (\bibinfo {year} {2018})}\BibitemShut {NoStop}%
\bibitem [{\citenamefont {Bongarzone}\ and\ \citenamefont {Gallaire}(2024)}]{bongarzone2024stick}%
  \BibitemOpen
  \bibfield  {author} {\bibinfo {author} {\bibfnamefont {A.}~\bibnamefont {Bongarzone}}\ and\ \bibinfo {author} {\bibfnamefont {F.}~\bibnamefont {Gallaire}},\ }\bibfield  {title} {\bibinfo {title} {Stick-slip-to-stick transition of liquid oscillations in a {U}-shaped tube},\ }\href {https://doi.org/0.1103/PhysRevFluids.9.034401} {\bibfield  {journal} {\bibinfo  {journal} {Phys. Rev. Fluids}\ }\textbf {\bibinfo {volume} {9}},\ \bibinfo {pages} {034401} (\bibinfo {year} {2024})}\BibitemShut {NoStop}%
\bibitem [{\citenamefont {Viola}\ and\ \citenamefont {Gallaire}(2018)}]{viola2018theoretical}%
  \BibitemOpen
  \bibfield  {author} {\bibinfo {author} {\bibfnamefont {F.}~\bibnamefont {Viola}}\ and\ \bibinfo {author} {\bibfnamefont {F.}~\bibnamefont {Gallaire}},\ }\bibfield  {title} {\bibinfo {title} {Theoretical framework to analyze the combined effect of surface tension and viscosity on the damping rate of sloshing waves},\ }\href {https://doi.org/10.1103/PhysRevFluids.3.094801} {\bibfield  {journal} {\bibinfo  {journal} {Phys. Rev. Fluids}\ }\textbf {\bibinfo {volume} {3}},\ \bibinfo {pages} {094801} (\bibinfo {year} {2018})}\BibitemShut {NoStop}%
\bibitem [{\citenamefont {Monsalve}\ \emph {et~al.}(2022)\citenamefont {Monsalve}, \citenamefont {Maurel}, \citenamefont {Pagneux},\ and\ \citenamefont {Petitjeans}}]{monsalve2022space-time}%
  \BibitemOpen
  \bibfield  {author} {\bibinfo {author} {\bibfnamefont {E.}~\bibnamefont {Monsalve}}, \bibinfo {author} {\bibfnamefont {A.}~\bibnamefont {Maurel}}, \bibinfo {author} {\bibfnamefont {V.}~\bibnamefont {Pagneux}},\ and\ \bibinfo {author} {\bibfnamefont {P.}~\bibnamefont {Petitjeans}},\ }\bibfield  {title} {\bibinfo {title} {Space-time-resolved measurements of the effect of pinned contact line on the dispersion relation of water waves},\ }\href {https://doi.org/10.1103/PhysRevFluids.7.014802} {\bibfield  {journal} {\bibinfo  {journal} {Phys. Rev. Fluids}\ }\textbf {\bibinfo {volume} {7}},\ \bibinfo {pages} {014802} (\bibinfo {year} {2022})}\BibitemShut {NoStop}%
\bibitem [{\citenamefont {Zhang}\ \emph {et~al.}(2023)\citenamefont {Zhang}, \citenamefont {Borthwick},\ and\ \citenamefont {Lin}}]{zhang2023pattern}%
  \BibitemOpen
  \bibfield  {author} {\bibinfo {author} {\bibfnamefont {S.}~\bibnamefont {Zhang}}, \bibinfo {author} {\bibfnamefont {A.~G.~L.}\ \bibnamefont {Borthwick}},\ and\ \bibinfo {author} {\bibfnamefont {Z.}~\bibnamefont {Lin}},\ }\bibfield  {title} {\bibinfo {title} {Pattern evolution and modal decomposition of {Faraday} waves in a brimful cylinder},\ }\href {https://doi.org/10.1017/jfm.2023.838} {\bibfield  {journal} {\bibinfo  {journal} {J. Fluid Mech.}\ }\textbf {\bibinfo {volume} {974}},\ \bibinfo {pages} {A56} (\bibinfo {year} {2023})}\BibitemShut {NoStop}%
\bibitem [{\citenamefont {Chandrasekhar}(1981)}]{chandrasekhar1981hydrodynamic}%
  \BibitemOpen
  \bibfield  {author} {\bibinfo {author} {\bibfnamefont {S.}~\bibnamefont {Chandrasekhar}},\ }\href@noop {} {\emph {\bibinfo {title} {Hydrodynamic and Hydromagnetic Stability}}}\ (\bibinfo  {publisher} {Dover Publications},\ \bibinfo {year} {1981})\BibitemShut {NoStop}%
\bibitem [{\citenamefont {Iyer}\ \emph {et~al.}(2020)\citenamefont {Iyer}, \citenamefont {Scheel}, \citenamefont {Schumacher},\ and\ \citenamefont {Sreenivasan}}]{iyer2020classical}%
  \BibitemOpen
  \bibfield  {author} {\bibinfo {author} {\bibfnamefont {K.~P.}\ \bibnamefont {Iyer}}, \bibinfo {author} {\bibfnamefont {J.~D.}\ \bibnamefont {Scheel}}, \bibinfo {author} {\bibfnamefont {J.}~\bibnamefont {Schumacher}},\ and\ \bibinfo {author} {\bibfnamefont {K.~R.}\ \bibnamefont {Sreenivasan}},\ }\bibfield  {title} {\bibinfo {title} {Classical 1/3 scaling of convection holds up to {Ra} = 10\textsuperscript{15}},\ }\href {https://doi.org/10.1073/pnas.1922794117} {\bibfield  {journal} {\bibinfo  {journal} {Proc. Natl. Acad. Sci. U.S.A.}\ }\textbf {\bibinfo {volume} {117}},\ \bibinfo {pages} {7594} (\bibinfo {year} {2020})}\BibitemShut {NoStop}%
\bibitem [{\citenamefont {Pelgrom}(2012)}]{dig_noise_marcel}%
  \BibitemOpen
  \bibfield  {author} {\bibinfo {author} {\bibfnamefont {M.~J.~M.}\ \bibnamefont {Pelgrom}},\ }\href@noop {} {\emph {\bibinfo {title} {Analog-to-digital conversion}}}\ (\bibinfo  {publisher} {Springer},\ \bibinfo {year} {2012})\BibitemShut {NoStop}%
\bibitem [{\citenamefont {{\v S}van{\v c}ara}\ \emph {et~al.}(2024)\citenamefont {{\v S}van{\v c}ara}, \citenamefont {Smaniotto}, \citenamefont {Solidoro}, \citenamefont {MacDonald}, \citenamefont {Patrick}, \citenamefont {Gregory}, \citenamefont {Barenghi},\ and\ \citenamefont {Weinfurtner}}]{svancara2024rotating}%
  \BibitemOpen
  \bibfield  {author} {\bibinfo {author} {\bibfnamefont {P.}~\bibnamefont {{\v S}van{\v c}ara}}, \bibinfo {author} {\bibfnamefont {P.}~\bibnamefont {Smaniotto}}, \bibinfo {author} {\bibfnamefont {L.}~\bibnamefont {Solidoro}}, \bibinfo {author} {\bibfnamefont {J.~F.}\ \bibnamefont {MacDonald}}, \bibinfo {author} {\bibfnamefont {S.}~\bibnamefont {Patrick}}, \bibinfo {author} {\bibfnamefont {R.}~\bibnamefont {Gregory}}, \bibinfo {author} {\bibfnamefont {C.~F.}\ \bibnamefont {Barenghi}},\ and\ \bibinfo {author} {\bibfnamefont {S.}~\bibnamefont {Weinfurtner}},\ }\bibfield  {title} {\bibinfo {title} {Rotating curved spacetime signatures from a giant quantum vortex},\ }\href {https://doi.org/10.1038/s41586-024-07176-8} {\bibfield  {journal} {\bibinfo  {journal} {Nature}\ }\textbf {\bibinfo {volume} {628}},\ \bibinfo {pages} {66–70} (\bibinfo {year} {2024})}\BibitemShut {NoStop}%
\end{thebibliography}%

\clearpage\raggedbottom\newpage

\interlinepenalty=0
\onecolumngrid

\begin{center}
    \Large\bfseries{Supplemental document}
\end{center}

\setcounter{section}{0}
\renewcommand{\theequation}{S\arabic{equation}}
\setcounter{equation}{0}
\renewcommand{\thefigure}{S\arabic{figure}}
\setcounter{figure}{0}

\section{Photocurrent due to a displaced laser beam}

\begin{figure}[htbp]
    \centering
    \includegraphics[scale=1]{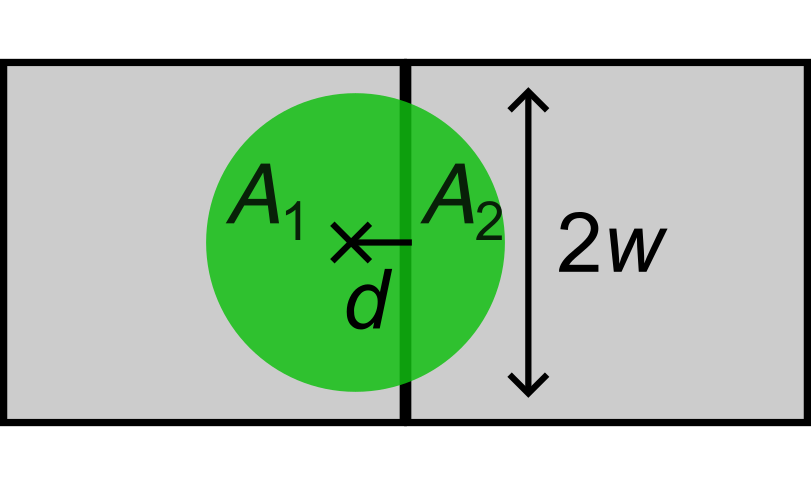}
    \caption{Laser beam of waist radius $w$ unequally illuminates two elements of a photodiode due to its displacement $d$ with respect to the photodiode's centre.}
    \label{fig:s1}
\end{figure}

Consider a laser beam with with the $1/e^2$ intensity radius $w$ illuminating a dual-element photodiode (DEPD), as shown in Fig.~\ref{fig:s1}. Since the beam is displaced by $d$ off the photodiode's centre, the laser illuminates unequal areas $A_1$ and $A_2$, respectively. Assuming homogeneous power density of the beam, $I = P/(A_1 + A_2)$, where $P$ is the laser power, the photocurrents induced in each element scale with the illuminated area and the photodiode's responsivity, $s = \eta e/(h \nu)$, where $\eta$ is the quantum efficiency, $e$ the electron charge, $h$ the Planck constant, and $\nu$ the light frequency. It therefore follows, for photocurrents $i_1$ and $i_2$, that
\begin{equation}
    i_1 = s A_1 I\,, \qquad    i_2 = s A_2 I\,.
    \label{eq:i1}
\end{equation}
Assuming that $d \ll w$, the illuminated areas can be expressed as
\begin{equation}
    i_1 = s \left[\frac{\pi w^2}{2} + 2wd\right] I\,, \qquad
    i_2 = s \left[\frac{\pi w^2}{2} - 2wd\right] I\,.
\end{equation}
We are interested in the photocurrent difference, $i_1 - i_2$. By normalising this quantity with the total photocurrent $i_1 + i_2$, we obtain
\begin{equation}
    \frac{i_1-i_2}{i_1+i_2} = \frac{s \left(\dfrac{\pi w^2}{2} + 2wd\right) I - s \left(\dfrac{\pi w^2}{2} - 2wd\right) I}{s \left(\dfrac{\pi w^2}{2} + 2wd\right) I + s \left(\dfrac{\pi w^2}{2} - 2wd\right) I} = \frac{4d}{\pi w}\,.
    \label{eq:idiff}
\end{equation}
As specified in the main text, the beam displacement is related to the local inclination angle $\theta$ as $d = 2F\theta$, where $F$ denotes the objective's focal length. By substituting the expression for $d$ into \eqref{eq:idiff} and rearranging, we obtain
\begin{equation}
    \theta = \frac{d}{2F} = \frac{\pi w}{8F} \frac{i_1 - i_2}{i_1 + i_2} =  \frac{\pi \mathrm{NA}}{8} \frac{i_1-i_2}{i_1+i_2}\,,
\end{equation}
where $\mathrm{NA}=w/F$ denotes the numerical aperture. This equation appears as Eq.~(2) in the main text, where the photocurrents are replaced by the corresponding voltage signals $V_1$ and $V_2$. In the experiment, each photocurrent is proportionally converted into an amplified voltage along the readout line, which is then digitised and stored (cf. Fig.~1 of the main text or Fig.~\ref{fig:s2}).

\section{DEPD offsetting procedure}

The two photosensitive elements of the DEPD are separated by a small gap, which reduces the detector's sensitivity when the laser beam is centred ($d \approx 0$) and its displacement fluctuates on a scale comparable to the gap width. To circumvent this detrimental effect, we employ an offsetting procedure described below. This approach also mitigates the effect of frequency doubling in the inclination spectra, which arises when the differential photocurrent periodically changes sign due to alternating illumination of the two DEPD elements.

\begin{figure}[htbp]
    \centering
    \includegraphics[scale=1]{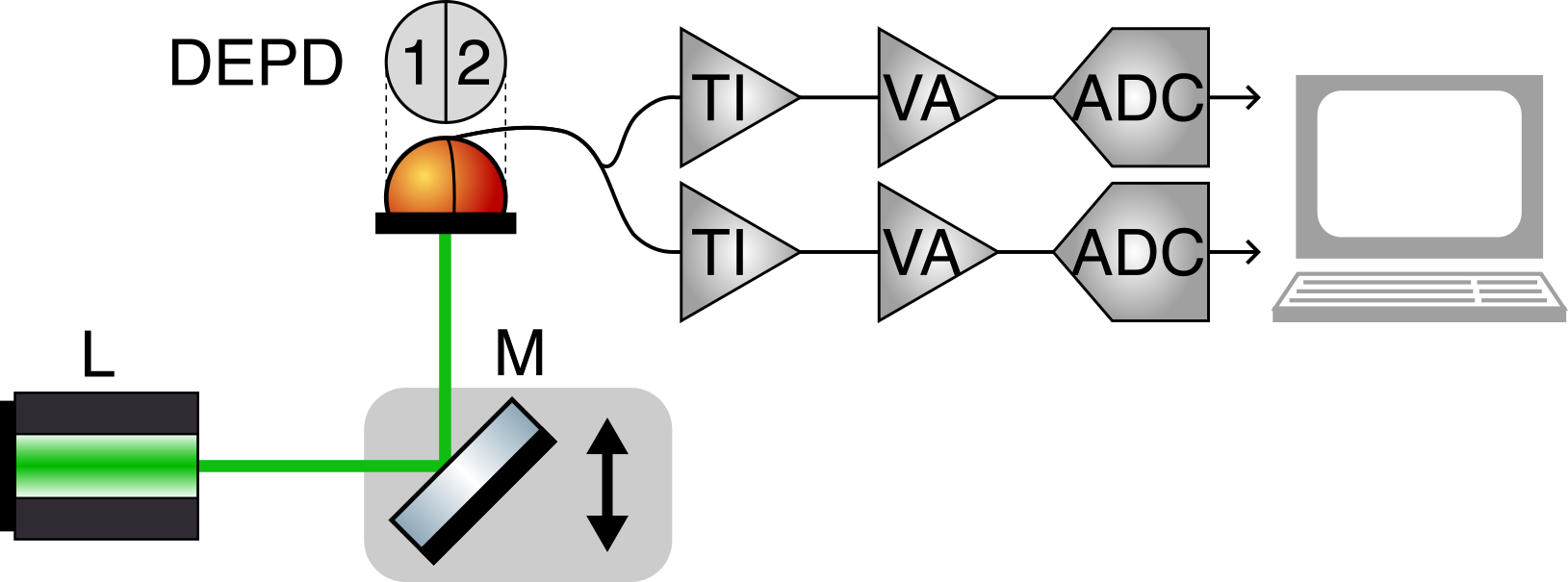}
    \caption{Scheme of the set-up for the offsetting procedure. L, laser; M, mirror on a piezo-actuated translation stage; DEPD, dual-element photodiode; TI, transimpedance amplifier; VA, voltage amplifier; ADC, analog-to-digital converter.}
    \label{fig:s2}
\end{figure}

The set-up for the offsetting procedure is outlined in Fig.~\ref{fig:s2}. The laser beam (L) is reflected off a mirror (M) mounted on a piezo-actuated translation stage, allowing us to precisely control the beam's displacement with respect to the DEPD. The resulting photocurrent difference is readout using the same acquisition line as described in the main text, consisting of a transimpedance amplifier (TIA), voltage amplifier (VA) and an analog-to-digital converter (ADC).

\begin{figure}[hbtp]
    \centering
    \includegraphics[scale=1]{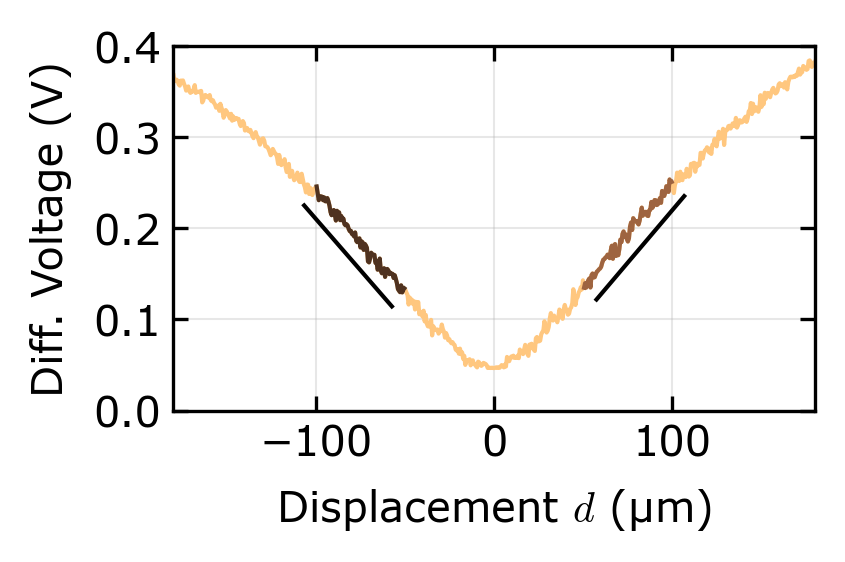}
    \caption{Response of the DEPD sensor to a controlled beam displacement. Brown traces highlight regions where the absolute differential signal scales linearly with the beam's displacement. Black lines are linear fits of these regions, vertically offset for clarity.}
    \label{fig:s3}
\end{figure}

A typical response curve, obtained by sweeping the laser beam across the DEPD, is shown in Fig.~\ref{fig:s3}. Large absolute differential voltages at the sweep extremities correspond to cases where the beam illuminates only one DEPD element, while the central minimum indicates that both elements are illuminated equally. Two approximately linear regions, highlighted in each side of the response curve, mark the range suitable for detecting interface waves. This linear response is consistent with \eqref{eq:idiff}, making it applicable also to the off-centre configuration.

To probe interface fluctuations, we impose the same lateral offset (approx. $75~\mu$m from the DEPD centre) during the alignment of the full set-up. As noted in the main text, this causes the differential signal to oscillate around a non-zero mean, which is subsequently removed during post-processing.

\section{Thermal stability of the fluid sample}

\begin{figure}[htbp]
    \centering
    \includegraphics[scale=1]{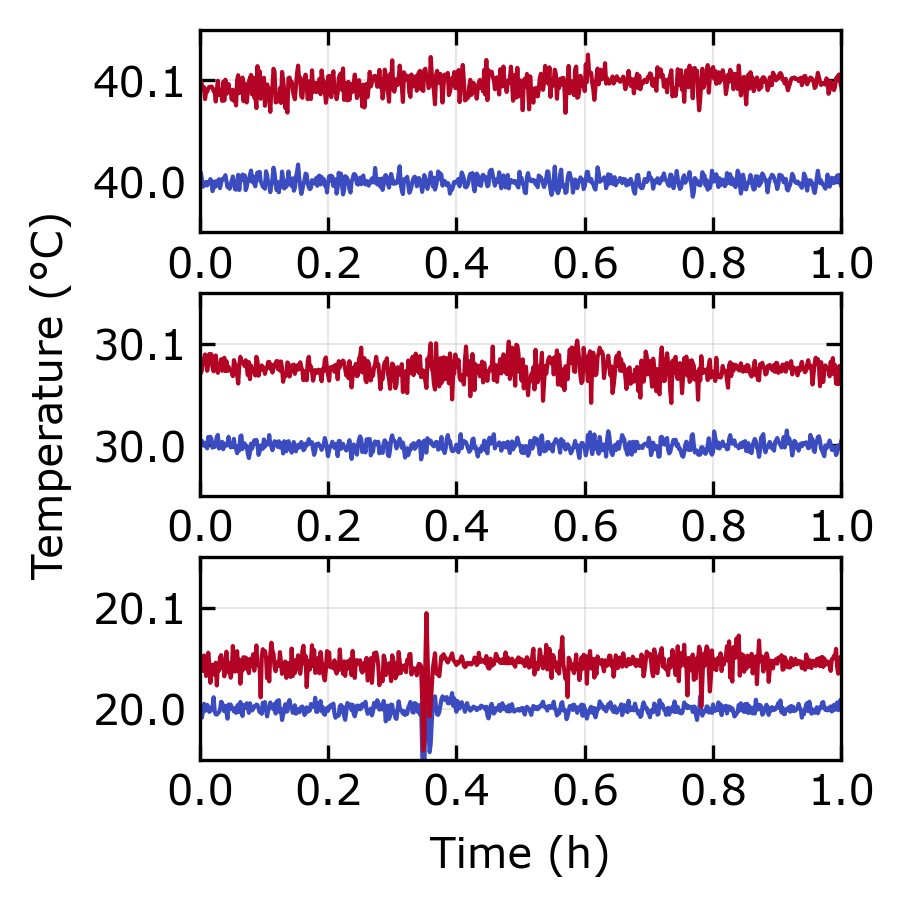}
    \caption{Time traces of two Pt-1000 thermometers monitoring the temperature of the sample cell. Blue lines: thermometer placed $2$~mm under the interface, and used as the feedback sensor for the PID controller. Red lines: thermometer placed $7$~mm under the interface.}
    \label{fig:s4}
\end{figure}

To regulate and maintain a constant temperature of the fluid sample, a Peltier element thermally connects the experimental cell to a copper heat sink located beneath it (cf. Fig.~2 in the main text). Depending on the polarity of the supplied current, the Peltier device can either heat or cool the sample. In our experiments, we operate it exclusively in heating mode, which allows for more precise temperature stabilisation. Fine control of the temperature is achieved by regulating the Peltier current using a software-based proportional-integral-derivative (PID) controller. Two Pt-1000 thermometers are embedded flush with the sample cell wall, positioned at distances of 2~mm and 7~mm from the fluid interface. The closer sensor provides feedback for the PID loop, while the second one monitors thermal gradients within the sample. Using this set-up, we achieved stable temperature control in the range of $20$-$40^\circ$C. Fig.~\ref{fig:s4} shows representative time traces of the measured temperature, demonstrating thermal stability over one hour intervals. The standard deviation of temperature measured by the control sensor (blue traces) is approximately 5~mK across the entire temperature range, while the auxiliary sensor (red traces) exhibits fluctuations of around 10~mK.

The two thermometers consistently measure a small temperature difference, which is partly due to their unequal calibration. We estimate this systematic offset of approx. 44~mK from the bottom panel of Fig.~\ref{fig:s4}, where the temperature is stabilised near the ambient laboratory temperature, suggesting that the sample cell is effectively in thermal equilibrium with its surroundings. At elevated set points, higher temperature differences are observed, as shown in the upper panels of Fig.~\ref{fig:s4}. This additional temperature difference reflects heat transport between the bottom of the sample cell and the liquid-air interface. The nature of thermal transport within the fluid is characterised by the Rayleigh number,
\begin{equation}
    \mathrm{Ra} = \frac{\rho\beta\Delta T l^3g}{\mu\alpha}\,,
    \label{eq:ra}
\end{equation}
where $\rho$ is the fluid density, $\beta$ the thermal expansion coefficient, $\Delta T$ the temperature difference across the cell's height $l = h_0 =  12$~mm, $g$ the gravitational acceleration, $\mu$ the dynamic viscosity, and $\alpha$ the thermal diffusivity. After subtracting the systematic offset, the temperature difference of $30$~mK ($50$~mK) is measured over the distance of $5$~mm for the $30^\circ$C ($40^\circ$C) case. Interpolating the temperature gradient over the entire height of the sample cell and using standard material properties of water, we obtain from \eqref{eq:ra} $\mathrm{Ra} \approx 3,100$ ($\mathrm{Ra} \approx 7,900$). Both these values are \emph{above} the critical Rayleigh number $\mathrm{Ra}_c \approx 1,708$ \cite{chandrasekhar1981hydrodynamic} that marks the onset of convective heat transfer. We therefore conclude that thermal convection might be an additional effect, partially exciting low-frequency waves presented in Fig.~6 of the main text. However, we do not expect the convective flow to transition into turbulence, which occurs for $\mathrm{Ra}$ values exceeding $\mathrm{Ra}_c$ by many orders of magnitude \cite{iyer2020classical}.

\section{Noise estimation}
The readout line outlined in Fig.~1 of the main text consists of a dual-element photodiode, transimpedance amplifier (conversion ratio $R_f = 10$~kV/A), voltage amplifier (gain $G = 10$), and analog-to-digital converter. Instead of DEPD photocurrents $i_{1,2}$, we acquire voltages $V_{1,2} = R_f G i_{1,2} + V_\mathrm{noise}$, where the additional noise is collected along the readout line. Typically, the time-averaged, total signal, $V_0 = \left\langle V_1 + V_2\right\rangle \approx 0.3$~V, which indicates that the total photocurrent induced by the laser beam on the DEPD is approximately $i_0 = V_0/(R_f G) \approx 3~\mu$A.

The first noise we consider is the photodiode shot noise. Assuming that each DEPD element outputs, on average, current $i_0/2$, the shot noise power spectrum, expressed as the resulting voltage signal acquired at the end of our readout line, reads
\begin{equation}
    S_\mathrm{SN} = 2e \frac{i_0}{2} \left(R_f G\right)^2 = e V_0 R_f G = 4.8\times 10^{-15}~\mathrm{V^2/Hz}\,,
\end{equation}
where $e$ is the elementary charge. The most important noise of the transimpedance amplifier is the input current noise, specified by the manufacturer as $i_\mathrm{TIA} = 5~\mathrm{pA}/\sqrt{\mathrm{Hz}}$. The resulting spectral density can be estimated as
\begin{equation}
    S_\mathrm{TIA} = \left(i_\mathrm{TIA} R_f G\right)^2 = 2.5\times 10^{-13}~\mathrm{V^2/Hz}\,.
\end{equation}
The output noise of the voltage amplifier is specified by the manufacturer as $V_\mathrm{VA} = 9~\mathrm{nV}/\sqrt{\mathrm{Hz}}$, i.e. the corresponding spectral density is
\begin{equation}
    S_\mathrm{VA} = V_\mathrm{VA}^2 = 8.1\times 10^{-17}~\mathrm{V^2/Hz}\,.
\end{equation}
Finally, we consider the quantisation noise of the analog-to-digital converter, which is given as \cite{dig_noise_marcel}
\begin{equation}
    S_\mathrm{ADC} = \frac{V_r^2}{6\times 2^{2b} f_s} = 2.0 \times 10^{-14}~\mathrm{V^2/Hz}\,,
\end{equation}
where $V_r = 0.5$~V is the ADC input range and $f_s = 2$~MHz is the sampling rate of the converter. Although the converter operates with 14-bit values, its effective bit depth is $b = 10$.

By combining these contributions, we obtain the total noise spectral density
\begin{equation}
    S_\mathrm{tot} = \left(S_\mathrm{SN}^2 + S_\mathrm{TIA}^2 + S_\mathrm{VA}^2 + S_\mathrm{ADC}^2\right)^{1/2} = 2.5\times 10^{-13}~\mathrm{V^2/Hz}\,.
\end{equation}
Note that $S_\mathrm{tot}$ is here limited by the input current noise of the transimpedance amplifier.

What remains now is to convert this value to the power spectral density of the inclination angle. Using standard error propagation techniques, we use Eq.~(2) of the main text, and obtain
\begin{equation}
    S_\theta = \left(\frac{\pi \mathrm{NA}}{8}\right)^2 \frac{\sqrt{2}S_\mathrm{tot}}{V_0^2} = 9.8\times 10^{-17}~\mathrm{rad^2/Hz}\,.
    \label{eq:stheta}
\end{equation}
We note in passing that in Fig.~4 of the main text, the noise floor increases by a factor of $2.7$ for the largest numerical aperture. These elevated noise levels exceed the signal amplitudes presented -- for example, in Fig.~3, where the spectrum plateaus at approximately $2 \times 10^{-17}~\mathrm{rad^2/Hz}$. Nonetheless, the thermal shape of the inclination spectra remains clearly observable across the entire frequency range. This suggests that the estimate given by \eqref{eq:stheta} is conservative, and that the dominant noise contribution from the transimpedance amplifier is likely lower than the manufacturer's specified value.

We finally remark that the inclination spectra presented in the main text are calculated using Welch's method. This approach estimates the spectral density of a time series by splitting the signal into $2,000$ overlapping segments. The individual spectra calculated over these segments are then averaged. This averaging process reduces the variance of the spectral estimate, effectively lifting coherent spectral features above the background noise.

\section{Boundary conditions}

\begin{figure}[htbp]
    \centering
    \includegraphics[scale=1]{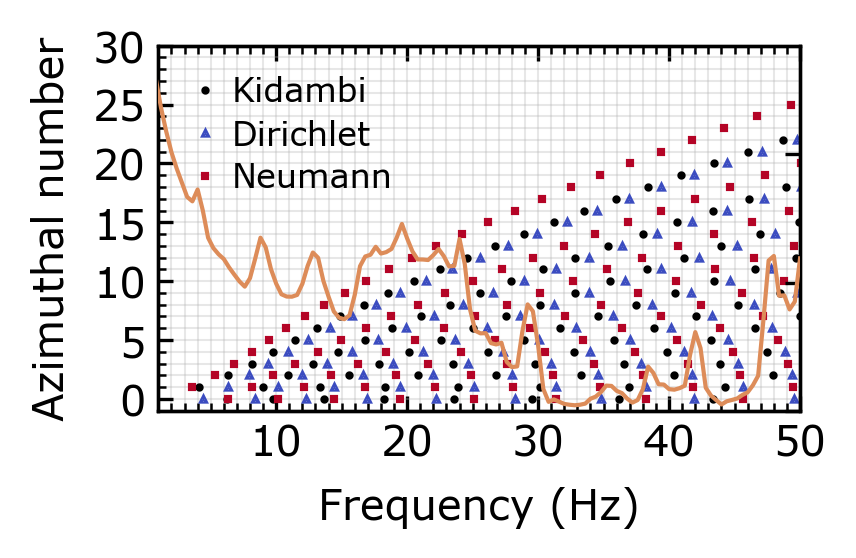}
    \caption{Frequencies of surface eigenmodes expected in the sample cell and satisfying the wetting boundary condition of Kidambi (black points), Dirichlet boundary condition (blue triangles) and Neumann boundary condition (red squares). The average spectrum of surface inclination, shown in Fig.~8 of the main text, is presented here as a background orange line for reference.}
    \label{fig:s5}
\end{figure}

To illustrate the rapid increase in the number of surface eigenmodes in our cell, Fig.~\ref{fig:s5} presents the predicted eigenfrequencies for two standard boundary conditions: Dirichlet (blue triangles) and Neumann (red squares). These conditions typically serve as first-order approximations for interface dynamics, with the notable exception of superfluid helium-4, where the Neumann condition has been shown to accurately reproduce experimental observations \cite{svancara2024rotating}. For comparison, we also include the numerical model developed by Kidambi \cite{kidambi2009capillary} (black points), which is used in Fig.~8 of the main text. The latter model yields eigenfrequencies that fall between the Dirichlet and Neumann predictions, reflecting a more realistic contact-line boundary condition in cylindrical geometry.

\end{document}